\documentclass[letterpaper,english,aps,prb,floatfix,twocolumn,amsfonts,superscriptaddress,longbibliography]{revtex4-1}

\usepackage[T1]{fontenc}
\usepackage[applemac]{inputenc}
\usepackage[english]{babel}
\usepackage{graphics}
\usepackage{amssymb}
\usepackage{amsmath}
\usepackage{overpic}
\usepackage{ulem}
\usepackage{natbib}

\usepackage{xcolor}
\usepackage{soul}

\newcommand{\be}{\begin{equation}}
\newcommand{\ee}{\end{equation}}
\newcommand{\bea}{\begin{eqnarray}}
\newcommand{\eea}{\end{eqnarray}}

\usepackage{dcolumn}
\usepackage{bm}
\usepackage{color}
\usepackage{float}

\def\a{\alpha}
\def\b{\beta}

\def\d{\delta}
\def\g{\gamma}

\def\O{\Omega}

\def\up{\uparrow}
\def\pll{\parallel}
\def\down{\downarrow}

\def\bb{{\bf b}}
\def\bk{{\bf k}}

\def\bq{{\bf q}}

\def\bA{{\bf A}}

\def\bR{{\bf R}}
\def\bS{{\bf S}}

\def\nn{\nonumber}
\def\lb{\label}
\def\pref#1{(\ref{#1})}

\newcount\bozza \bozza=0
\ifnum\bozza=1
\newdimen\shift \shift=-2truecm
\def\lb#1{%
{\label{#1}\rlap{\kern\shift{$\scriptstyle#1$}}}}
\else\def\lb#1{\label{#1}} \fi

\begin{document}
\title{Third harmonic generation from collective modes in disordered superconductors} 
\author{G.~Seibold} 
\affiliation{Institut f\"ur Physik, BTU Cottbus-Senftenberg, PBox 101344, 03013 Cottbus,
Germany}
\author{M.~Udina}
\affiliation{Department of Physics and ISC-CNR, "Sapienza" University of Rome, P.le Aldo Moro 5, 00185, Rome, Italy}
\author{C.~Castellani}
\affiliation{Department of Physics and ISC-CNR, "Sapienza" University of Rome, P.le Aldo Moro 5, 00185, Rome, Italy}
\author{L.~Benfatto} 
\affiliation{Department of Physics and ISC-CNR, "Sapienza" University of Rome, P.le Aldo Moro 5, 00185, Rome, Italy}
 

\begin{abstract}
Recent experiments with strong THz fields in both conventional and unconventional superconductors have clearly evidenced a marked third-harmonic generation below the superconducting temperature $T_c$. Its  interpretation challenged substantial theoretical work aimed at establishing the relative efficiency of quasiparticle excitations and collective modes in triggering such a resonant response. Here we compute the non-linear current by implementing a time-dependent  Bogoljubov de-Gennes approach, with the twofold aim to account non-perturbatively for the effect of local disorder, and to include the contribution of all collective modes, i.e. superconducting amplitude (Higgs) and phase fluctuations, and charge fluctuations. We show that, in agreement with previous work, already at small disorder the quasiparticle response is dominated by paramagnetic effects. We further demonstrate that paramagnetic processes mediate also the response of all collective modes, with a substantial contribution of charge/phase fluctuations. These processes, which have been overlooked so far, turn out to dominate the third-order current at strong disorder. In addition, we show that disorder strongly influences the polarization dependence of the non-linear response, with a marked difference between the clean and the disordered case. Our results are particularly relevant for recent experiments in cuprates, whose band structure is in a first approximation reproduced by our lattice model. 
\end{abstract}

\maketitle

\section{Introduction}

In the last two decades the technological advances in the generation and manipulation of intense light pulses opened the way to a revolutionary tool for investigating complex materials\cite{giannetti_review}. As compared to ordinary spectroscopic methods, the use of short and intense pump pulses offers two main advantages. From one side, combining an external perturbation faster than the typical relaxation times of the system with a time-delayed weak probe has the potential to access a genuine non-equilibrium condition, hence disclosing physical phenomena not observable by standard spectroscopies. From the other side, the high impulsive value of the electromagnetic (e.m.) field triggers naturally a non-linear optical response, with selection rules in general complementary with respect to linear response. The latter aspect becomes predominant in the case of THz multicycle light pulses, whose duration is typically of several ps. In this case, the frequency spectrum of the pump pulses is rather narrow, so one has in principle the opportunity to detect in the field reflected or transmitted through the sample higher harmonics of the central frequency of the pump. 
This effect turns out to be particularly pronounced when the pumping process excites resonantly a specific propagating mode of the system. Indeed, the enhancement of the high-harmonic generation intensity can be used to identify the characteristic frequencies of the  excitations under scrutiny, and their dependence on external parameters like e.g.\ the temperature or the magnetic field.  

A successful example of non-linear THz driving of collective modes has been provided by the case of superconducting (SC) systems, where pronounced third-harmonic generation (THG) has been reported when the system is cooled  below the critical temperature $T_c$\cite{shimano_science14,shimano_prb17,shimano_prl18,wang_natphot2019,kaiser_natcomm20,shimano_prb20}. To understand how this effect can be used to characterize collective electronic modes, and to distinguish them from the particle-hole continuum, one can start considering the general expansion of the current in powers of the vector potential ${\bf A}(t)$, so that 
\be
\lb{jgen}
j_{\a}=\chi_{\a\b}^{(1)}A_\b+\chi_{\a\b\g\d}^{(3)}A_\b A_\g A_\d,
\ee
where the convolution in time has been omitted for simplicity, and $\a$ denotes the spatial components. The THG is linked to the properties of the $\chi_{\a\b\g\d}^{(3)}$ non-linear optical kernel, which can be computed starting from a given microscopic interacting model in the presence of a gauge field $\bA$. Even though the primary coupling of $\bA$ is to the fermionic particle-hole excitations, these can mediate the decay to an intermediate collective mode $\Phi$, which is then resonantly driven by the e.m. field. In particular, dropping space indeces for simplicity, if the energy expanded in the fluctuation field $\delta \Phi$ acquires a term scaling as $\gamma \delta\Phi A^2$, where $\gamma$ if an effective coupling,  the non-linear kernel acquires a term proportional to the collective-mode fluctuations $\langle \delta \Phi^2\rangle$, i.e. 
\be
\lb{chi3gen}
\chi^{(3)}(\omega)\sim \gamma^2 \langle \delta \Phi^2\rangle (\omega),
\ee
with $\omega$ frequency of an ideal monochromatic pump field. By combining Eq.s\ \pref{jgen} and \pref{chi3gen} it is easy to show that the intensity of the THG scales as:
\be
\lb{ithg}
I^{THG}\sim  |\chi^{(3)}(2\omega)|^2.
\ee
Assuming that the spectrum of $\Phi$ is resonant near a characteristic frequency $\omega_\Phi$, as e.g.\ $\langle \delta \Phi^2\rangle(\omega)\sim 1/(\omega^2-\omega_\Phi^2)$, one then obtains an enhancement of the THG \pref{ithg} when the pump frequency matches half the value of the mode resonance, $\omega=\omega_{\Phi}/2$. When the experiment is carried out with a pump-probe setup, one can show\cite{giorgianni_natphys19,udina_prb19} that the resonance of the non-linear kernel \pref{chi3gen} manifests as oscillation of the differential probe field at the characteristic frequency $\omega_\Phi$, as a function of the pump-probe delay time $t_{pp}$. The theoretical understanding of the non-linear kernel resonances requires then to solve two separate problems: (i) the identification of all the collective modes $\Phi$ which exist besides the particle-hole continuum, and their spectrum near the corresponding resonance frequency $\omega_\Phi$; (ii) the computation of their effective non-linear coupling to light $\gamma$, and its tensorial structure. Indeed, by changing the polarization axis of $\bA$ with respect to the crystallographic axes one can access experimentally the various components of the $\chi^{(3)}$ tensor. 

In the case of superconductors, the transition occurs thanks to the spontaneous  breaking of the continuous $U(1)$ gauge symmetry. The equilibrium value of the order parameter controls the gap $\Delta$  in the single-particle excitation spectrum. In clean systems this leads to a well-defined structure of the quasiparticle continuum, which only appears at low temperatures above the threshold frequency $\omega=2\Delta$.  This effect is manifest in the BCS response function describing density fluctuations at long wavelength 
($\bq=0)$ and finite frequency. However, density fluctuations induced by an uniform potential should vanish, so this result is a well-known demonstration that the BCS approximation violates charge conservation \cite{schrieffer,deveraux_review,cea_raman_prb16,chubukov_prb17}, that can only be restored adding the contribution of charge  and phase fluctuations beyond BCS. Nonetheless, the  non-linear optical response has a leading contribution controlled by lattice-modulated charge fluctuations, which are not constrained to any conservation rule, as it has been usually discussed within the context of the Raman response of superconductors\cite{deveraux_review,cea_raman_prb16,chubukov_prb17}. As a consequence, below $T_c$ the modification of the quasiparticle spectrum  appears also in the THG  with a sharp enhancement  at $2\omega=2\Delta$\cite{cea_prb16}. 

In addition,  two collective SC modes emerge below $T_c$, connected to the amplitude and phase of the complex order parameter\cite{varma_review,shimano_review19}. The phase mode is the Goldstone excitation connecting energetically equivalent ground states of the system, and the corresponding phase rigidity against non-uniform phase fluctuations is connected to the superfluid density\cite{nagaosa}. The phase fluctuations are conjugated to the density ones, whose spectrum is dominated in a charged system by plasma excitations, usually occurring at hundreds of THz. The amplitude mode, also named Higgs mode for the analogy with the massive Higgs boson of the Standard model, has a characteristic mass frequency of the same order of the quasiparticle threshold, i.e.\ $\omega_H=2\Delta$, which is about few THz in conventional superconductors\cite{varma_review,shimano_review19}. In these systems it has been proven experimentally by means of both pump-probe protocols\cite{carbone_pnas12,shimano_prl13,shimano_science14} and THG measurements\cite{shimano_science14,shimano_prb17,wang_natphot2019} that the non-linear optical kernel presents a marked resonance at $\omega_\Phi=2\Delta$. Since this coincides with the Higgs-mode frequency, these results have been initially naturally linked to fluctuations of the Higgs mode\cite{shimano_science14,aoki_prb15}. However, subsequent work\cite{cea_prb16} highlighted the importance of the BCS response, which is resonant at the same $2\Delta$ frequency and is much larger than the Higgs response, since for clean superconductors one finds that $\chi^{(3)}_{BCS}\gg \chi^{(3)}_{Higgs}$. More specifically, this results follows from the fact that  the Higgs mode is weakly coupled to light, $\gamma_{Higgs}\simeq 0$, so the overall intensity of the Higgs signal is too small to be detected. On the other hand, the analysis of Ref.\ \cite{cea_prb16} for a two-dimensional square lattice showed also that the two signals can be distinguished by their tensorial structure. Indeed, while the Higgs response is fully isotropic, the BCS one is strongly anisotropic. By decomposing $\chi^{(3)}$ in terms of the irreducible representation of the square lattice, $\chi^{(3)}_{Higgs}$ has $A_{1g}$ symmetry, while $\chi^{(3)}_{BCS}$ has $B_{1g}$ symmetry, so the THG is expected to almost vanish when the field is applied along the diagonal of the square lattice\cite{cea_prb16}. Interestingly, subsequent analysis\cite{shimano_prb17}  of the polarization dependence of the THG in NbNshowed a marked $A_{1g}$  signal, challenging again the interpretation of the experimental results. Indeed, by using a microscopic multiband SC model appropriate for NbN, both $\chi^{(3)}_{Higgs}$ and $\chi^{(3)}_{BCS}$ have a sizeable non-symmetric component\cite{cea_prb18}. 

A considerable step forward in the theoretical interpretation of the THG in conventional superconductors has been provided by the inclusion of disorder effects\cite{silaev_prb19,shimano_prb19}. Indeed, these works have shown that point-like impurities can reverse the order of the Higgs and BCS contributions, making $\chi^{(3)}_{Higgs}\gg \chi^{(3)}_{BCS}$ by increasing disorder.  Within a minimal-couping approach $\bA$ couples microscopically to the fermionic particle-hole excitations, via two terms named diamagnetic and paramagnetic. The former one is a coupling of $\bA^2$ to the inverse mass tensors, which reduces to $n/m$ for a parabolic band dispersion, with $n$ electron density and $m$ electron mass. The latter is a coupling of $\bA$ to the electronic current. Because of current conservation, in clean superconductors paramagnetic terms are completely ineffective to mediate a finite coupling of the Higgs to the light, so $\gamma_{Higgs}\simeq \gamma_{dia}$, and $\gamma_{dia}\simeq 0$ as mentioned above. However, in the presence of even small disorder the paramagnetic processes become finite and the resulting $\gamma_{Higgs}\simeq \gamma_{para}\gg \gamma_{dia}$, triggering a strong response of the Higgs mode\cite{silaev_prb19,shimano_prb19,shimano_review19}. 

While these results can help understanding some of the experimental findings, several open issues  still need to be addressed. So far, the analysis of Ref.\ \cite{silaev_prb19,shimano_prb19} focused only on the effects of disorder on the Higgs-light coupling, neglecting completely the role of SC phase and charge fluctuations. This is however dangerous within continuum models as the ones considered in Ref.\ \cite{silaev_prb19,shimano_prb19}, where the inclusion of  phase/charge  fluctuations is crucial in order to preserve the gauge invariance of the theory\cite{schrieffer}, as indeed discussed within the context of Raman spectroscopy\cite{deveraux_review,cea_raman_prb16,chubukov_prb17}. In addition, as disorder increases the mixing between the amplitude, phase and charge fluctuations becomes unavoidable\cite{cea_prl15,trivedi_prb20}. As a consequence, a crucial open question is the nature of the full non-linear optical response in a disordered superconductor, obtained by including also phase and charge fluctuations. 
A further issue concerns the possibility to treat disorder exactly, going beyond the  approximated schemes employed so far\cite{silaev_prb19,shimano_prb19}. For example, the approaches used in Ref.\ \cite{silaev_prb19,shimano_prb19}  provide different analytical estimates for the disorder-induced modifications of $\chi^{(3)}_{Higgs}$ and $\chi^{(3)}_{BCS}$ as a function of the dimensionless quantity $2\Delta \tau$, $\tau$ being the quasiparticle scattering rate. Finally, a crucial issue for the comparison with the experiments is the evaluation of the polarization dependence of  $\chi^{(3)}_{Higgs}$ and $\chi^{(3)}_{BCS}$ tensors with the inclusion of disorder. Once more, this question cannot be answered within the simplified continuum model considered in Ref.s\ \cite{silaev_prb19,shimano_prb19}, since it requires to introduce a proper lattice structure. A recent\cite{tsuji_cm20} theoretical calculation done using a lattice model appropriate for NbN suggests indeed a considerable change of the polarization dependence of the non-linear optical kernel $\chi^{(3)}$ as disorder is included. 

In this work we investigate the THG response for a disordered superconducting system by including amplitude, phase and charge collective modes. We use a time-domain approach where we compute the non-linear current  within the time-dependent Bogoljubov-de Gennes (BdG) theory. As compared to previous work, where disorder has been included in the self-consistent Born approximation\cite{silaev_prb19,tsuji_cm20} or by implementing an approximate Mattis-Bardeen limit\cite{shimano_prb19}, our method has the advantage to treat disorder {\it exactly}, and to include all possible self-energy and vertex corrections to the kernel describing the non-linear response within BCS. This approach has been shown to be crucial in the intermediate- and strong-disorder regime\cite{trivedi_prb01,dubi_nature07,cea_prb14,cea_prl15,seibold_prb17,trivedi_prb20}, where a non perturbative treatment of disorder is required to account for the emergent inhomogeneity of the SC background, that has a direct impact on the collective-mode behavior. We focus on the same two-dimensional lattice structure considered in the clean case in Ref.\ \cite{cea_prb16}, since it can be relevant for recent experiments in cuprate superconductors\cite{shimano_prl18,kaiser_natcomm20,shimano_prb20}. Our approach confirms that disorder enhances BCS paramagnetic contributions to the current, and it simoultaneously triggers a sizeable contribution from {\it all} collective modes. By considering only Higgs fluctuations we recover the results of previous work, i.e.\ that in the strong-disorder regime the Higgs contribution overcomes the BCS one. However, when adding charge and phase fluctuations we find that their contribution is immediately present even for very small disorder, and it becomes the dominant one at large disorder. This result is directly linked to what observed in previous work within the context of the optical conductivity of strongly-disordered superconductors, where it has been shown that disorder triggers a finite paramagnetic sub-gap absorption due to charge and phase modes\cite{cea_prb14,seibold_prb17,pracht_prb17}. In addition, we analyze the polarization dependence of the THG, showing that the BCS contribution changes radically from the clean to the dirty case. In the homogenous case, where diamagnetic processes rule out the THG, the response is maximum for a field applied along the crystallographic directions, and it almost vanishes for a field applied along the diagonal, as already found in previous work by means of the perturbative expansion\cite{cea_prb16}. Conversely, in the presence of even weak disorder the paramagnetic BCS processes have opposite anisotropy, with a shallow minimum for a field applied along the crystallographic axis, while all collective modes give rise to an almost isotropic contribution. As we shall discuss below, these results are particularly relevant for recent experiments in cuprate superconductors\cite{shimano_prl18,shimano_prb20,kaiser_natcomm20}.

The paper is organized as follows: in Sec. \ref{sec:form}  we outline
the model and we present the time-dependent Bogoljubov-de Gennes  formalism from which
we extract the nonlinear dynamics. In Sec. \ref{sec:res} we first check the formalism to compute the linear response in the presence of disorder, that will be also used to characterize the disorder level via the quasiparticle transport scattering time $\tau$. In Sec.\ \ref{sec:THG} we compute the third-order response, first in the clean case, where we recover the results obtained previously by means of a perturbative diagrammatic expansion, and then in the disordered case, for several disorder levels. In Sec. \ref{sec:dis} we present the results at various disorder levels  and for different field ploarization.  In Sec. \ref{sec:conc} we discuss our outcomes on the light of previous theoretical and experimental work, and we present a summary of the results. The technical details are reported in various appendices. In Appendix \ref{appHom} we show in details the equivalence, in the homogeneous case, between our time-dependent approach and previous theoretical approaches based on the so-called Anderson pseudospin dynamics or on the effective-action formalism. Appendix \ref{appNum} contains additional technical details on the numerical procedure used to extract the third-order current. In Appendix \ref{appcurr} we show explicitly the subleading contributions to the third-order currents triggered by disorder. 
Finally in Appendix \ref{appTau} we detail the procedure used to estimate the transport scattering rate from the optical conductivity.

\section{Formalism}\label{sec:form}

\subsection{Definition of the third-order current}
Previous work for clean systems has shown that the THG including collective modes can be easily computed in frequency space by means of a  diagrammatic expansion\cite{cea_prb16,aoki_prb15,shimano_prb17}, that is fully equivalent to a perturbative solution by using the so-called pseudospin dynamics\cite{shimano_science14,aoki_prb15}. Within the same scheme in the presence of disorder one should also include self-energy and vertex corrections due to disorder\cite{silaev_prb19,shimano_prb19,tsuji_cm20}, and resort to some approximation to compute the response. Here we employ a different procedure, and we compute the non-linear current in the time domain by using the time evolution of the density matrix, where all the collective modes can be included. The components of the density matrix, which correspond to normal and anomalous average of fermionic operators in the SC state, are expressed by means of the elements of the Bogoljubov-de Gennes (BdG) transformation, which in turn includes the effects of disorder. By averaging over several disorder configurations we are able to treat disorder at a non-perturbative level. As a consequence, while SC properties are still treated at mean-field level, the disorder is treated exactly. This implies in principle to include all possible self-energy and vertex corrections to the fermionic susceptibilities controlling the non-linear response, allowing us to provide results ranging from the weak-disorder limit, where the self-consistent Born approximation used in Ref.\ \cite{silaev_prb19,tsuji_cm20} is expected to hold, to the strong-disorder limit, implemented in an approximated way in Ref.\ \cite{shimano_prb19}.

As a starting point we consider the attractive Hubbard model on the two-dimensional square lattice, with local
disorder                                                 
\begin{equation}\label{eq:hub}                                         
H=-t\sum_{ij\sigma}c^\dagger_{i\sigma}c_{j\sigma} - |U|\sum_{i}n_{i\uparrow}n_
{i\downarrow} +\sum_{i\sigma}V_i n_{i\sigma}                                      
\end{equation}
where the local potential $V_i$ is taken from a flat distribution
$-V_0 \le V_i \le +V_0$. Only the nearest neighbor hopping $ t$
is considered so that the parameters $|U|/t$ and $V_0/t$ specify the
energy scales of the problem. As already discussed in the clean case in Ref.\ \cite{cea_prb16}, this SC model  represents  a good prototype model to study  the polarization dependence of the THG induced by the lattice structure, as encoded in the electronic band dispersion. In addition, its disordered version \pref{eq:hub} has been already proven\cite{trivedi_prb01,dubi_nature07,cea_prb14,seibold_prb17,trivedi_prb20} to contain the main ingredients needed to discuss the effects of disorder in the equilibrium optical response, as we shall discuss below. 

Eq.\ (\ref{eq:hub}) admits a  SC ground state, where the SC order parameter is defined as $\Delta_n= -|U| \langle c_{n,\downarrow} c_{n,\uparrow}\rangle$. The resulting mean-field quadratic Hamiltonian can be diagonalized by using the BdG  transformation
\be
\lb{cidef}
c_{i\sigma}=\sum_k\left[u_i(k)\gamma_{k,\sigma}-\sigma v_i^*(k)\gamma_{k,-\sigma
}^\dagger\right],
\ee
which yields the eigenvalue equations
\begin{align}
\omega_k u_n(k)&=\sum_{j}t_{nj} u_j(k) + [V_n-\frac{|U|}{2}\langle n_n\rangle -
\mu] u_n(k) \nonumber \\ 
 \lb{eq1} 
&+\Delta_n v_n(k),                              \\                                                                                                                                     
\omega_k v_n(k)&=-\sum_{j}t^*_{nj} v_j(k) -
[V_n-\frac{|U|}{2}\langle n_n\rangle -\mu] u_n(k)\nonumber \\
\lb{eq2}     
&+\Delta^*_n u_n(k).
\end{align}
From the eigenvalue problem Eqs. \pref{eq1}-\pref{eq2}) one can
iteratively determine the ground state density matrix
${\cal R}$ with the elements:

\begin{eqnarray*}
  \rho_{ij}&=&\langle c_{i,\uparrow}^\dagger c_{j,\uparrow} \rangle \\
  &=&\sum_k\left\lbrack v_i(k)v_j^*(k)(1-f(E_k))+u_i^*(k)u_j(k)f(E_k)\right\rbrack, \\
\bar{\rho}_{ij}&=&  \langle c_{i,\downarrow} c_{j,\downarrow}^\dagger \rangle \\
&=&\sum_k\left\lbrack u_i(k)u_j^*(k)(1-f(E_k))+v_i^*(k)v_j(k)f(E_k)\right\rbrack, \\
\kappa_{ij}&=& \langle c_{i,\downarrow} c_{j,\uparrow} \rangle\\
&=&\sum_k\left\lbrack -u_i(k)v_j^*(k)(1-f(E_k))+v_i^*(k)u_j(k)f(E_k)\right\rbrack.
\end{eqnarray*}
Here $f(E)$ is the Fermi function, which reduces to a step function in the  $T=0$ limit considered below. In a compact notation the density matrix can be written as
\begin{equation}\label{eq:dm}
  {\cal R}=\left(
  \begin{array}{cc}
    \rho & \kappa^\dagger \\
    \kappa & \bar{\rho}
  \end{array} \right) \,. 
\end{equation}  
The BdG energy can then be expressed via the elements of the density matrix as
\begin{eqnarray*}
  E^{BdG}&=-t&\sum_{ij}\left(\rho_{ij}-\bar{\rho}_{ij}\right) +
  -|U|\sum_{i}\left( \rho_{ii} (1-\bar{\rho}_{ii})+\kappa^*_{ii}\kappa_{ii}\right)\\
&+&\sum_{i}V_i\left\lbrack \rho_{ii}-\bar{\rho}_{ii}+1 \right\rbrack,
\end{eqnarray*}                                                                  
and the elements of the BdG-hamiltonian are defined as
\begin{equation}\label{eq:bdg}
  {\cal H}^{BdG}_{ij}=\frac{\partial E^{BdG}}{\partial {\cal R}_{ji}} \,.
\end{equation}

Finally, the dynamics of the density matrix can be computed from
\begin{equation}\label{eq:mot}
  i\frac{d}{dt}{\cal R}=\left\lbrack {\cal R}, {\cal H}^{BdG}\right\rbrack\,.
\end{equation}

In the absence of an external field the density matrix ${\cal R}$ and the Hamiltonian ${\cal H}^{BdG}$ commute, so it simply follows from Eq.\ (\ref{eq:mot}) that  the density matrix has no time evolution. The dynamics of ${\cal R}(t)$ is induced via the coupling to the electromagnetic field $\vec{E}(t)=-\partial \vec{A}(t)/\partial t $. Let us first consider the case of a (spatially constant) field along the $x$ direction. $A_x(t)$ is coupled to the system via the Peierls
substitution $c_{i+x,\sigma}^\dagger c_{i,\sigma} \rightarrow
e^{i A_x(t)} c_{i+x,\sigma}^\dagger c_{i,\sigma}$, where for simplicity we will drop form the equations all the constant by putting the lattice spacing, the electronic charge $e$, the light velocity $c$ and the Planck constant $\hbar$ equal to one. Since we are interested in computing the current up to the third order in the gauge field, see Eq.\ \pref{jgen}, we need to  expand the Peierls phase
factor in the Hamiltonian up to fourth order:
\begin{eqnarray}
  H^{BdG}&=&H^{BdG}_0-\sum_{n}j_p(R_n)A_x(t) \label{eq:coup} \\
  &-&\frac{1}{2}\sum_n j_{dia}(R_n) A_x^2(t) \nonumber\\
  &+& \frac{1}{6}\sum_n j_p(R_n)A_x^3(R_n) +\frac{1}{24}\sum_n j_{dia}(R_n)A_x^4(t) \,,\nonumber
\end{eqnarray}
  with para- and diamagnetic currents defined as
  \begin{eqnarray}
    j_{para}(R_n) \label{eq:jpara} 
    &=& i t \left \lbrack \rho_{i+x,i}-\bar{\rho}_{i,i+x} - h.c.\right\rbrack \\
    j_{dia}(R_n) \label{eq:jdia} 
&=&    -t \left \lbrack \rho_{i+x,i}-\bar{\rho}_{i,i+x} + h.c.\right\rbrack,
  \end{eqnarray}
  where the explicit time dependence of the density-matrix elements in Eq.\ \pref{eq:jpara}-\pref{eq:jdia} has been omitted. 
The diamagnetic term $j_{dia}(R_n)$ is different from zero also when $A_x=0$, in which case it coincides with
 the kinetic energy $t_x(R_n)$ on the
  bond between sites $R_n$ and $R_{n+x}$. In a continuum model the last two terms of the  expansion \pref{eq:coup} are absent, since the minimal-coupling substitution only admits linear and quadratic terms in the gauge field. In Eq.s\ \pref{eq:jpara}-\pref{eq:jdia} the paramagnetic and diamagnetic currents  depend explicitly on the gauge field via the  elements of the density matrix 
$\rho$ and $\bar\rho$.  The total resulting
current up to third-order in $A_x$ can be obtained from the derivative of Eq.\ \pref{eq:coup} with respect to $A_x$, so that 
\begin{eqnarray}
  j(t)&=&-\frac{1}{N}\frac{\partial {\cal H}^{BdG}}{\partial A_x}  \nn\\
  \label{eq:jfexp}
  &=&(1-\frac{1}{2}A_x^2)j_{para}(A)+(A_x-\frac{1}{6}A_x^3)j_{dia}(A) \end{eqnarray}
where we introduced the total para- and diamagnetic currents
$j_{para/dia}(A)=1/N \sum_n j_{para/dia}(R_n)$, and we made explicit their dependence on time via the applied vector potential $A_x(t)$. Here $N$ denotes the number of lattice sites.

To build a power expansion in the applied gauge field we will consider a monochromatic case where $A_x(t)=A_0 \sin(\Omega t)$. To obtain $j$ at a given order in $A_0$ we should consider the explicit dependence on $A_x$ in Eq.\ \pref{eq:jfexp}, along with a power expansion of the paramagnetic and diamagnetic currents as a power series in $A_0$
\begin{eqnarray}
  j_{para/dia}(t)&=&j_{para/dia}^{(0)} +A_0 j_{para/dia}^{(1)}(t)\label{eq:jpexp}  \\&+&A_0^2 j_{para/dia}^{(2)}(t)+ A_0^3 j_{para/dia}^{(3)}(t) + \dots \nonumber
\end{eqnarray}
with $j_{para}^{(0)}=0$ and by definition the Fourier components of $j_{para/dia}^{(m)}$
are independent on $A_0$. Notice that since $A_0$ has to be taken as a small number, i.e.\ $A_0 \sim 10^{-3}$, it is not affordable to read off the third-harmonic term $|j(3\Omega)|$ from the Fourier transformation of
  $j(t)$. In contrast, the various terms $j_{para/dia}^{(m)}$ can be determined by computing $j(t)$ for different values of the prefactor $A_0$, as detailed in Appendix B.

The linear response limit of Eq.\ (\ref{eq:jfexp}) is obtained from:
\be
\lb{jlin}
 j^{1th}(t)=A_0j_{para}^{(1)}+A_0 T_x^0,
 \ee
 where we used the fact that $j_{dia}^{(0)}\equiv T_x^0 $, and $T_x^0 =1/N \sum_n t_x(R_n)$ is the 
total kinetic energy along the $x$-direction, with the naught superscript indicating the evaluation in the BCS ground state. In a continuum model, this would reduce to the familiar $n/m$ diamagnetic term. In linear-response theory one can also write $A_0j^{(1)}_{para}(t)=\int dt' \chi_{jj}(t-t')A_x(t')$ where  $\chi_{jj}(t-t')=-i\langle{\cal T}j_p(t),j_p(t')\rangle$ is the current-current correlation function within the standard Kubo formalism. As we shall see in the next Section, the resulting optical conductivity obtained from the present time-dependent approach coincides with the corresponding result   obtained previously \cite{seibold_prb17,cea_prb14} within the Kubo formalism.

\begin{figure}[hhh]
  \includegraphics[width=8.5cm,clip=true]{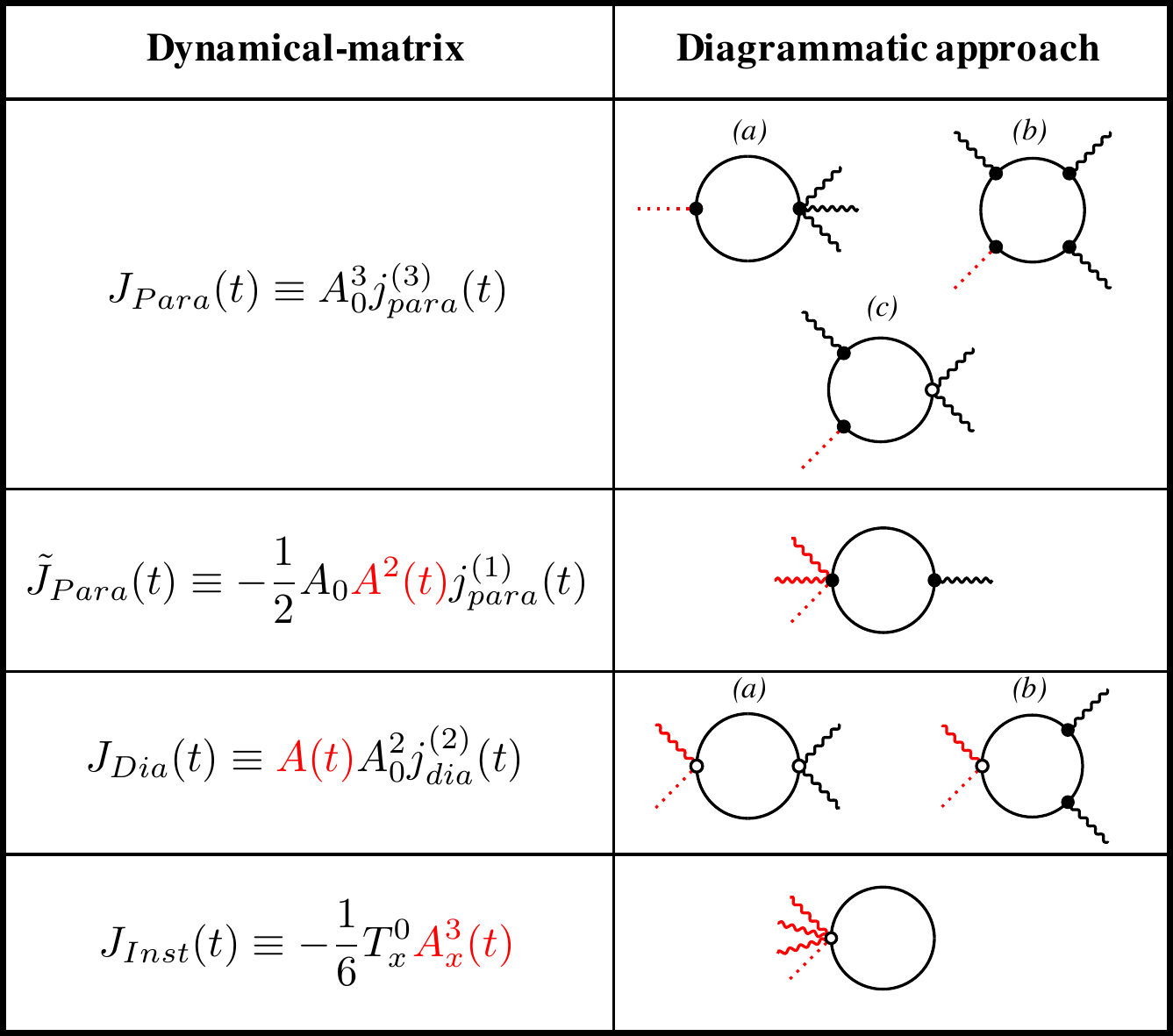}
  \caption{Classification of the BCS response to the third-order current in terms of the diagrammatic representation used in previous work\cite{cea_prb16,shimano_prb17,cea_prb18,silaev_prb19,shimano_prb19,tsuji_cm20}. Here solid black lines denote the  Green's function of electrons. In our time-domain approach each contribution to the third-order current has an explicit dependence on the field, that we highlighted in red both in the definitions (left column) and in the diagrammatic representation (red wavy lines in Feynman's diagrams). The remaining field dependence comes from the response of $j^{(m)}_{para/dia}$ to the field, which is denoted with black wavy lines.  In part of the previous work the diagrammatic representation has been used for the quartic action in the gauge field, whose derivative with respect to $A$ defines the current\cite{cea_prb16,cea_prb18,shimano_prb17,tsuji_cm20}. To help to establish the analogies with the same processes discussed before, we then denoted with a dotted red line the $\bA$ field with respect to which the derivative has been performed in Eq.\ \pref{eq:jfexp} to compute the current. In the presence of disorder all self-energy and vertex corrections due to disorder (not shown) are automatically included. } 
\label{diag}                                                   
\end{figure}  
   
For the 3rd harmonic contribution to the current Eq.\  (\ref{eq:jfexp})  gives:
\begin{eqnarray}
  j^{3rd}(t)&=&A_0^3 j_{para}^{(3)}(t)
  -\frac{1}{2}A_0 A_x^2(t)j_{para}^{(1)}(t)\nn\\
  \label{eq:j3rd}
  &+& A_x(t) A_0^2 j_{dia}^{(2)}(t)
  -\frac{1}{6}T_x^0 A_x^3(t).
  \end{eqnarray}
As discussed above, the second and last term are only present in a lattice model, since they originate from the last two terms of the expansion \pref{eq:coup}.

The present approach allows us to investigate the influence of collective modes on $j^{3rd}(t)$  by selectively including the corresponding dynamics in the equations of motion Eq.\ (\ref{eq:mot}). In fact, the BdG hamiltonian Eq.\ (\ref{eq:bdg}) is a function of the local density $\rho_{nn}(t)$ and the Gorkov function $\kappa_{nn}(t)=|\kappa_{nn}(t)| e^{i\varphi_n(t)}$,
which have to be computed at each instant of time. The time dependence of these quantities corresponds to the dynamics of charge ($\rho_{nn}(t)$), amplitude ($|\kappa_{nn}(t)|$) and phase ($\varphi_n(t)$) modes, which affect the paramagnetic and diamagnetic currents \pref{eq:jpara}-\pref{eq:jdia}.  The BCS approximation, i.e.\ the neglect of collective
excitations, corresponds instead to the evaluation of ${\cal R}$ by keeping
the initial values (at $t=t_0$) of the local density and the Gorkov function
in the BdG hamiltonian Eq.\ (\ref{eq:bdg}), i.e.\ by setting
$\rho_{nn}(t)=\rho^{(0)}_{nn}(t_0)$ and $\kappa_{nn}(t)=\kappa^{(0)}_{nn}(t_0)$.
Similarly, one can selectively include only the amplitude modes by setting
$\rho_{nn}(t)=\rho^{(0)}_{nn}(t_0)$ and $\varphi_{n}(t)=\varphi^{(0)}_{n}(t_0)$,
but keeping the dynamics of $|\kappa_{nn}(t)|$. Below, we also
investigate the effect of including only charge and phase modes, by setting
$\kappa_{nn}(t)=\kappa^{(0)}_{nn}(t_0)$ while keeping the dynamics
of $\rho_{nn}(t)$ and $\varphi_n(t)$.

\subsection{Contributions to the third-order current}\label{3rd:def}
\begin{figure*}[htb]
 \includegraphics[width=15cm,clip=true]{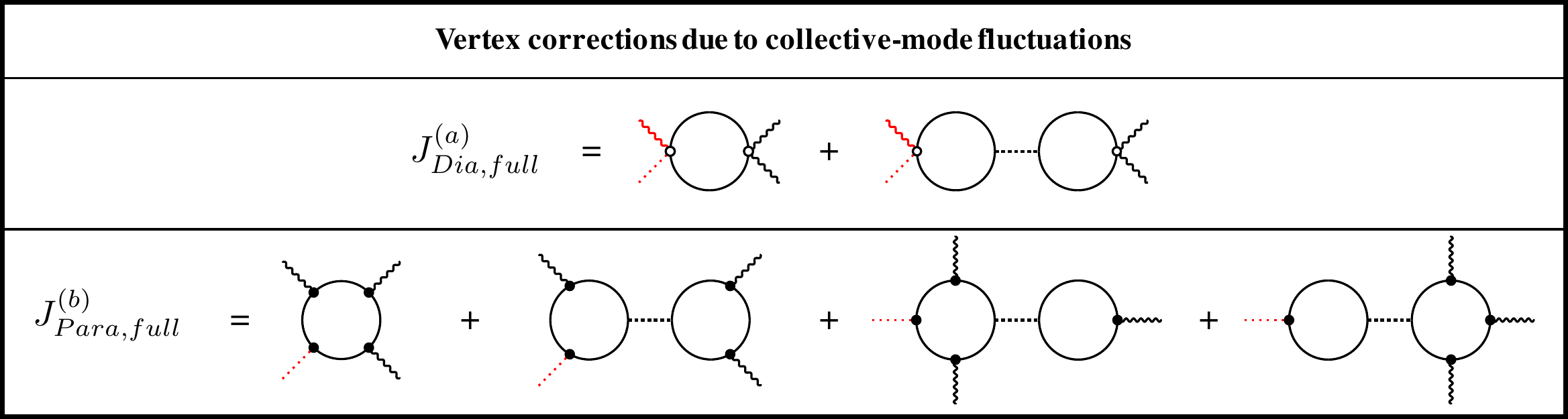}
  \caption{Vertex corrections due to collective-mode fluctuations to selected third-order processes shown in Fig.\ \ref{diag}. Top panel: vertex corrections to $J_{Dia}^{(a)}$, that is the only process present in the clean limit. Left: bare-bubble contribution. Right: vertex correction due to a given collective mode (Higgs/phase/charge), where the dashed line denotes the RPA resummation of the potential in the corresponding amplitude/phase/charge sector. In the presence of disorder the three modes become mixed. Bottom panel: vertex corrections to $J_{Para}^{(b)}$, that would be zero in the clean limit. Left: bare-bubble contribution. Right: vertex corrections, with the same notation as before. }
  \label{diag2}                                                   
\end{figure*}

In the absence of disorder it is easy to recover the equivalence between the dynamical-matrix equations of motions \pref{eq:mot} and the standard perturbative approach discussed in previous work, within either the so-called  Anderson pseudospin approach\cite{shimano_science14,aoki_prb15}, or the effective-action formalism\cite{cea_prb16,cea_prb18,udina_prb19}. 
A non-perturbative solution has been also recently discussed in Ref.s \cite{wu_prb19,wang_prb20} within either a kinetic equation or density-matrix approach  in relation to the observation of second-harmonic generation  in a superconductor in the presence of a finite dc component of the THz field\cite{wang_natphot2019,wang_prl20}. 
As shown in details in Appendix A, the equations of motion \pref{eq:mot} are fully equivalent to the equations of motions of the so-called Anderson pseudospins\cite{shimano_science14,aoki_prb15}, that just represent a specific combinations of the density-matrix components.  For the homogeneous case the equations can be solved explicitly in momentum and frequency space, to obtain the coefficients of the expansion of the current \pref{eq:jfexp} in powers of the gauge field, see Eq.\ \pref{jgen}.  Such an expansion is the same obtained in Ref.\ \cite{cea_prb16, cea_prb18,udina_prb19} by means of an effective-action formalism, where after integrating out the fermionic degrees of freedom the coefficients of the effective action for the gauge field correspond to fermionic loops. These can be represented by means of Feynman diagrams, with "bare" terms representing the BCS response in the absence of collective-mode fluctuations, and "dressed" diagrams which include vertex corrections due to collective modes computed at RPA level. To make a closer connection between the two approaches, we report  in Fig.\ \ref{diag}  the Feynman diagrams corresponding to the various terms of Eq.\ \pref{eq:j3rd} in the absence of collective modes, by denoting with a dotted red line the field with respect to which the derivative has been performed in Eq.\ \pref{eq:jfexp}. Diagrams with only two vertexes will be denoted as "Kubo" contributions, since they can be expressed as correlations functions either for the current, $\chi_{jj}$, or for the kinetic energy, $\chi_{kk}$, where 
\be
\lb{corr}
\chi_{jj/kk}(t-t')=-i\langle{\cal T}j_{para/dia}(t),j_{para/dia}(t')\rangle.
\ee
The logical construction of the terms is always the same. A contribution $j_{para/dia}^{(m)}$ is obtained by considering the response of $j_{para/dia}$ as given by Eq.\ \pref{eq:jpara}-\pref{eq:jdia} to a combinations of terms from the Hamiltonian \pref{eq:coup} leading to a power $A^m$. To distinguish a paramagnetic-like from a diamagnetic-like coupling between the gauge field and the fermions we use full/empty circles in Fig.\ \ref{diag}. The various vertexes are connected by electronic Green's functions, and the average over disorder guarantees that all diagrams are dressed by self-energy and vertex corrections due to disorder. The various contributions to Eq.\ (\ref{eq:j3rd}) can then be classified as follows:

\begin{itemize}
\item $A_0^3 j_{para}^{(3)}(t)$. We have three contributions, labeled as $a, b, c$  in Fig.\ \ref{diag}. The first one is a Kubo-like contribution that can be written as $\frac{1}{6}\int\!\!dt'\chi_{jj}(t-t')A^3(t')$, and it is obtained as the response of $j_{para}$
  to the  term $\frac{1}{6}\sum_n j_{para}(R_n)A_x^3(R_n)$ in the Hamiltonian (\ref{eq:coup}). This term is absent in a continuum model. We will denote this contribution $J_{Para}$ in what follows. 
    
    \item $-\frac{1}{2}A_0 A^2(t)j_{para}^{(1)}(t)$. This term contains only a Kubo-like contribution, i.e.\ $\frac{1}{2}A^2(t)\int\!\!dt'\chi_{jj}(t-t')A(t')$, which arises from   the response of the third term of  Eq.\ \pref{eq:jfexp} to the  linear term
  $-\sum_{n}j_{para}(R_n)A_x(R_n)$ in the Hamiltonian \pref{eq:coup}. This term, absent in the continuum model, is shown in Fig.\ \ref{diag}. We will denote it $\tilde J_{Para}$ in what follows. 
  
\item $A(t) A_0^2 j_{dia}^{(2)}(t)$. Here we have a Kubo-like contribution, $-\frac{1}{2}A(t)\int\!\!dt'\chi_{kk}(t-t')A^2(t')$, coming from
  the response of $j_{dia}$ to the term $-\frac{1}{2}\sum_n k_x(R_n) A_x^2(R_n)$ in the Hamiltonian, so it is determined by the kinetic-energy correlation function (see diagram labeled $a$ in Fig.\ \ref{diag}). This is the only Kubo-like contribution which is also present in a continuum model. A non-Kubo like term  is also present, diagram $b$ in Fig.\ \ref{diag}. We will denote this contribution $J_{Dia}$ in what follows.

\item  $ -\frac{1}{6}T_x^0 A_x^3(t)$. This term is again absent in a continuum model. As discussed in Ref.\ \cite{cea_prb18}, in the clean case such an instantaneous third-order response to the gauge field must be properly included in the lattice model to correctly capture the polarization dependence of the THG. We will denote this contribution $J_{Inst}$ in what follows.

\end{itemize}  

In summary, we will then use the following definitions to identify the various contributions to the third-order current:
\bea
J_{Para}&\equiv& A_0^3 j_{para}^{(3)}(t),\nn\\
\tilde J_{Para}&\equiv& -\frac{1}{2}A_0 A^2(t)j_{para}^{(1)}(t),\nn\\
J_{Dia}&\equiv& A(t) A_0^2 j_{dia}^{(2)}(t), \nn\\
\lb{jdeftot}
J_{Inst}&\equiv& -\frac{1}{6}T_x^0 A_x^3(t). \label{abbrev}
\eea

As mentioned above, whereas in the time-dependent approach collective modes are included via 
their time dependence in the corresponding elements of the dynamical matrix,
in the diagrammatic approach this is achieved by including all possible
vertex corrections due to amplitude, charge and phase fluctuations to the diagrams of Fig.\ \ref{diag}. In Fig.\ \ref{diag2} we show the response obtained by including SC fluctuations to, e.g., $J_{Dia}^{(a)}$ and $J_{Para}^{(b)}$.  The diamagnetic process $J_{Dia}^{(a)}$ and its collective-mode corrections are the only processes present in the clean limit, while disorder can trigger a paramagnetic contribution from all collective modes,  as  we will show explicitly in the next sections. As a consequence, the dashed line of Fig.\ \ref{diag2} represents fluctuations in all the amplitude, phase and charge channels, and also the relative mixing between the various modes, that becomes particularly importantly at strong disorder\cite{cea_prb14,cea_prl15,seibold_prb17,trivedi_prb20}. The advantage of the present approach based on the density matrix  is that all these effects can be automatically included by adding the dynamical evolution of the $\rho_{nn}(t)$ and $\kappa_{nn}(t)$ functions. Previous work\cite{silaev_prb19,shimano_prb19,tsuji_cm20} highlighted the importance of disorder to trigger the Higgs response. In the following we will show that disorder triggers also a non-trivial response from charge and phase modes, which at all disorder levels contribute  to the resonant THG at $\Omega=\Delta$.

A second advantage of the present method is that it allows us to study how disorder affects the polarization dependence of the THG, i.e.\ how the THG changes when the applied field $\bA$ is aligned along an arbitrary direction with respect to the main crystallographic axes. Its relevance to distinguish the various contributions has been already highlighted both in the clean\cite{cea_prb16,shimano_prb17,cea_prb18} and in the disordered\cite{tsuji_cm20} case. On very general grounds, for a field
applied at an angle $\theta$ with the $x$ direction the Fourier transform of the non-linear current has a simple decomposition  in a component  $j^{3rd}_\pll$ parallel to the field and one $j^{3rd}_\perp$, perpendicular to it,  with 
\bea
j^{3rd}_\pll(\theta)&=&\alpha-\beta \sin^2(2\theta)  \label{jpll},\\
j^{3rd}_\perp(\theta)&=&-(\beta/2) \sin(4\theta)\label{jperp},
\eea
where $\a,\b$ are functions of the field frequency $\Omega$. The decomposition \pref{jpll}-\pref{jperp} just relies on the fact that for the band structure we are considering, the third-order current along e.g.\ the $x$ direction can only scale as $\sim A_x^3$ or $\sim A_xA_y^2$, where the explicit dependence in frequency has been omitted. In the clean case, where only Kubo-like diamagnetic processes are relevant, the $\a,\beta$ functions can be further decomposed in correlation functions which distinguish the spatial-indexes permutations. 
Indeed, since for Kubo-like diagrams each vertex carries two components $A_i A_j$ of the field, one can express the spectral component of the third-order current along the $i-th$ direction as $j_i=\chi^{(3)}_{ij;km} A_jA_k A_m$, without any ambiguity in the definition of the tensor components. This further allows one to decompose the non-linear current using the irreducible representation of the $D_{4h}$ group for our lattice structure, so that
\bea
j^{3rd}_\pll(\theta)/A_0^3&=&\chi^{(3)}_{xx;xx}(\cos^4\theta+\sin^4\theta)+\nn\\
&+&2[\chi^{(3)}_{xx;yy}+\chi^{(3)}_{xy;xy}+\chi^{(3)}_{xy;yx}]\sin^2\theta\cos^2\theta=\nn\\
\lb{chidec}
&=&K_{A1g}+K_{B1g}\cos^2(2\theta)+K_{B2g}\sin^2(2\theta)
\eea
where we introduced the usual definitions:
\bea
\lb{ab}
K_{A1g/B1g}&=&\frac{\chi^{(3)}_{xx;xx}\pm \chi^{(3)}_{xx;yy}}{2},\\
K_{B2g}&=&\chi^{(3)}_{xy;xy}+\chi^{(3)}_{xy;yx}.
\eea
For the lattice structure used in the present paper only $\chi^{(3)}_{xx;xx}$ and $\chi^{(3)}_{xx;yy}$ are different from zero in the clean case, so that $K_{B2g}=0$.  By comparing Eq.s \eqref{chidec} and \eqref{jpll} one can also identify $K_{A1g}=\a-\beta$ and $K_{B1g}=\beta$. As discussed in Ref.\ \cite{cea_prb16}, in the absence of disorder $j^{3rd}(\theta=\pi/4)\propto K_{A1g}$ almost vanishes, since the THG response mediated by $J^{(a)}_{Dia}$ is almost purely $B_{1g}$, with a small $A_{1g}$ contribution due to the Higgs. 
However, in the presence of disorder the tensor components $\chi^{(3)}_{ij;km}$ becomes ill-defined, since also non-Kubo like diagrams as those contributing to $J_{Para}$ are present, see Fig.\ \ref{diag}.  Nonetheless, the third-order current still admits the general decomposition \pref{jpll}-\pref{jperp}. In what follows, for each disorder level and for each third-order contribution to the current we will compute the full frequency dependence of the current in the field direction $j^{3rd}_\pll$ for the two cases $\theta=0$ and $\theta=\pi/4$, in order to highlight the differences with respect to the clean case. In addition, we will compute the current at zero frequency by diagonalizing the BCS hamiltonian in the presence of a finite vector potential, similar to the approach used in Ref. \cite{seibold_prl12} for the evaluation of the superfluid stiffness in disordered systems.
This allows us to investigate the full angular dependence of the $j^{3rd}_\pll$ spectrum at $\Omega=0$, showing how the polarization dependence is affected by the presence of disorder. 

In the paper we present results for $|U|=2$, $n=0.875$ and various levels of disorder $V_0/t$ on $16\times 16$ lattices with periodic boundary conditions. In all the figures, the black dashed line will denote  the BCS results, i.e.\ the response in the absence of collective modes. The results obtained by including only the Higgs or the charge/phase fluctuations will be denoted with blue and green dashed lines, respectively. The red solid line will denote the full results including amplitude, phase and charge fluctuations. For each ploarization we will show the result of $j^{3rd}_\pll$, i.e.\ the third-order current in the field direction. Details of the numerical simulations implemented in the manuscript are given in Appendix B. 

\begin{figure}[hhh]
  \includegraphics[width=8cm,clip=true]{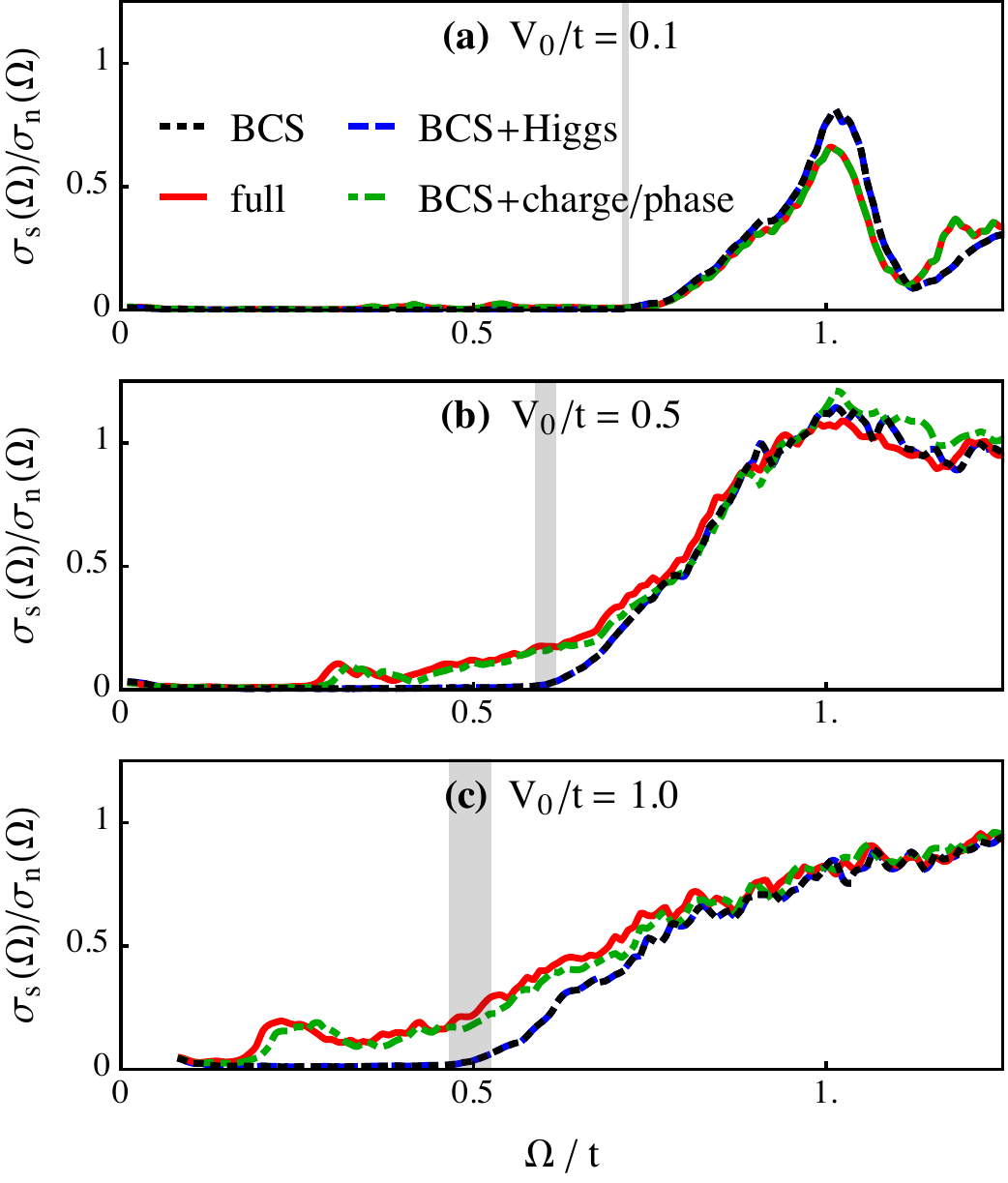}
  \caption{Optical conductivity $\sigma(\omega)$ for different degrees
    of disorder $V_0/t$, as labelled in the various panels. 
The black line denotes the BCS results, i.e.\ the response in the absence of collective modes. The red solid line is the full result including amplitude, phase and charge fluctuations, while blue and green lines denote the results obtained by including only the Higgs or the phase/charge fluctuations, respectively. Vertical gray bars denote the range of variation of the spectral gap. }
\label{opdc}                                                   
\end{figure}  

\section{Linear response and Optical conductivity}\label{sec:res}
To check the validity of the method, we start considering the linear current, as given by Eq.\ \pref{jlin}, and we extract from it the real part of the optical conductivity. The latter can be compared to the one obtained by means of the linear-response theory, whose implementation in the presence of disorder by including collective modes as been discussed in previous work\cite{cea_prb14,seibold_prb17}.  To make a closer connection with the standard Mattis-Bardeen (MB) theory\cite{MattisBardeen}, we plot in Fig.\ \ref{opdc} the optical conductivity $\sigma_s(\omega)$ normalized to its value $\sigma_n(\omega)$ in the non-SC state, obtained by setting the anomalous correlations $\kappa=0$ in Eq.\ \pref{eq:dm}.   

The BCS result (balck dashed line) for $\sigma(\omega)$, obtained without inclusion of collective modes,  follows the prediction of MB theory, with an absorption threshold located at the spectral gap $E_{gap}$,  which is indicated in Fig.\ \ref{opdc} by the vertical grey line. It is worth noting that in a clean superconductor $E_{gap}$ should just coincide with twice the SC order parameter $2\Delta$. For finite and weak disorder the
local order parameter $\Delta_n$ becomes spatially inhomogeneous, but  $E_{gap}$ still scales with twice the average order parameter, i.e.\ $E_{gap}\simeq 2\langle \Delta_n\rangle$. However, as discussed in previous work\cite{trivedi_prb01,dubi_nature07,cea_prb14,cea_prl15,seibold_prb17,trivedi_prb20}, as disorder increases further $E_{gap}$ is only slightly suppressed (eventually increasing again at strong disorder), while the average order parameter tends progressively to vanish, signalling a boson-like superconductor-insulator transition induced by the localization of Cooper pairs\cite{trivedi_prb01,dubi_nature07,cea_prb14,cea_prl15,seibold_prb17,trivedi_prb20}. For simplicity of notation,  in what follows we will  denote the spectral gap $E_{gap}=2\Delta$, since the energy scale which sets the enhancement of the THG is always half of the spectral gap. However, one should keep in mind that $\Delta$ coincides with the average order parameter only for weak or zero disorder. Due to the presence of a  van Hove singularity (vHs) in our band dispersion,  a pronounced absorption peak above $2\Delta$ appears for weak and intermediate disorder,  at the location of the DOS maximum, see Fig.\ \ref{opdc}a-b. This is consistent with the recent analysis of Ref.\ \cite{seibold_prb17}, where it has been shown how the analytical results from standard MB theory\cite{MattisBardeen}, formulated for a prototypical parabolic band dispersion, get slightly modified when the DOS significantly varies on the scale of the SC gap, as indeed is the case when the chemical potential is close to a vHs. When the collective-mode contribution is taken into account the results get modified, see red lines in Fig.\ \ref{opdc}, and a substantial subgap absorption appears below the $2\Delta$ edge. As discussed previously\cite{cea_prb14,seibold_prb17,pracht_prb17}, the extra absorption can be ascribed to charge and phase modes, which become optically
active for increasing disorder. This is evidenced by the dashed green line in Fig.\ \ref{opdc}, obtained by including only the dynamics of the phase and charge modes. Clearly, the corresponding result is very close to the full $\sigma(\omega)$ (red solid line), which includes also the coupling to the Higgs mode. The emergence of subgap absorption in strongly-disordered thins films has been observed experimentally in several systems\cite{armitage_prb07,driessen_prl12,dressel_natphys15,armitage_prb16,scheffler_prb16}, showing that a substantial deviation from the MB paradigm can also be used experimentally to estimate the level of disorder in the sample. 

The analysis of the linear response allows us to extract an effective scattering time $\tau$ for the electrons, that can be used to make a quantitative comparison with previous theoretical work\cite{silaev_prb19,shimano_prb19,tsuji_cm20} and with the experiments. The procedure is described in detail in Appendix \ref{appTau}. It is worth noting that our approach includes also Hartree corrections to the local chemical potential. Since the local electron density is itself inhomogeneous, this leads to an effective disorder potential that is larger than $V_0$. On the other hand, to make a comparison with the experiments what matters is only the final value of the scattering time $\tau$, as it manifests in observable quantities like the optical conductivity. In Appendix \ref{appTau} we then provide an estimate of $\tau$ for each value of $V_0$, based on a MB analysis of the computed optical conductivity. 



\section{Third harmonic response}\label{sec:THG}

\subsection{Clean system}\label{THGclean}

\begin{figure}[htb]
  \includegraphics[width=8cm,clip=true]{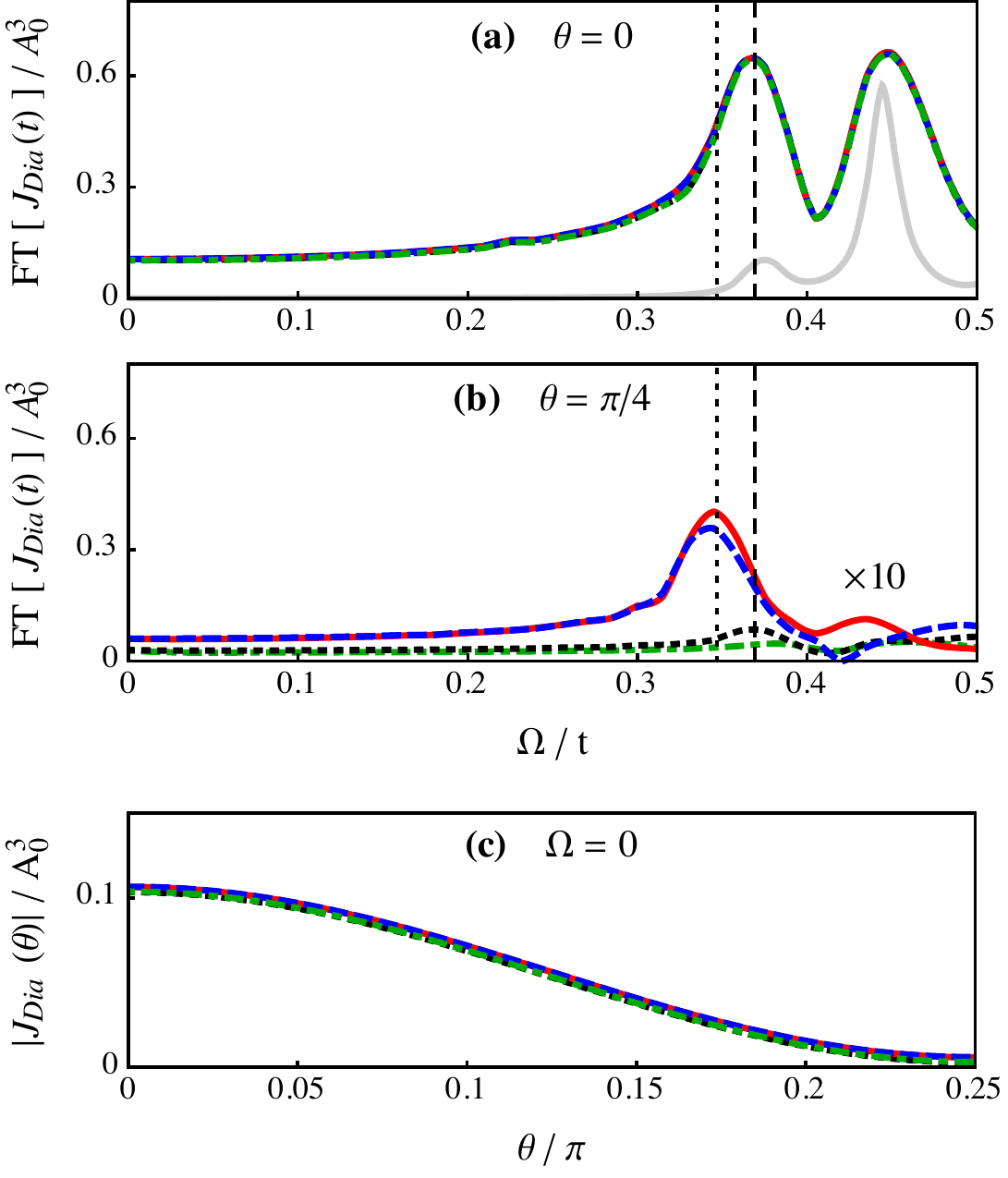}
  \caption{Spectrum of the diamagnetic contribution $J_{Dia}$ to the third-order current for a clean system at (a) $\theta=0$ (vector potential along the $x$ direction) and (b) $\theta=\pi/4$ (vector potential along the diagonal direction). The colour code for the black, blue, green and red lines is the same of Fig.\ \ref{opdc}. In panel (a) we also show for comparison the DOS $\rho(2\Omega)$ in the SC state (grey line), with a clear signature of the vHs at $\Omega/t=0.45$. The vertical dashed line indicates half of the spectral gap, while the vertical dotted line denotes the SC order parameter. The small difference between the two is due to finite-size effects. (c) Angular dependence of $J_{Dia}(\O=0)$ as a function of the polarization angle $\theta$.} 
\label{jdiav0}                                                   
\end{figure}

To check the validity of our approach we will first compare the results for the clean case, where the THG for the present band dispersion has been derived analytically in Ref.\ \cite{cea_prb16}, and reviewed in Appendix A.  In the clean case, the BCS response admits only the Kubo-like kinetic contribution $\chi_{kk}$ and the instantaneous response, given by diagrams $J^{(a)}_{Dia}$ and $J_{Inst}$   in Fig.\ \ref{diag}, respectively. Indeed, one can show that all diagrams with current-like coupling between the gauge field and the fermions vanish\cite{aoki_prb15,cea_prb16,cea_prb18,silaev_prb19}, as a consequence of current conservation. The $J_{Inst}$  contribution is constant in frequency and it is also present above $T_c$, so it cannot contribute to an enhancement of the THG in the SC state. Nonetheless, it should be included to correctly account for the polarization dependence in the clean case, as observed in Ref.\ \cite{cea_prb18}. As we shall see, in the presence of disorder it becomes rapidly subleading, so we only report its behavior for the sake of completeness in Appendix \pref{appcurr}. 
The $J^{(a)}_{Dia}$ diagram represents the Kubo response for lattice-modulated charge fluctuations, and it has a resonance at $\Omega=2\Delta$. As a consequence, since in computing the non-linear current each vertex carries an $A^2$ insertion, its contribution to the THG is maximum at $\Omega=\Delta$\cite{cea_prb16}. The same diamagnetic coupling mediates also the excitation of the Higgs mode, as shown by the second diagram in the top panel of Fig.\ \ref{diag2}. This process can be overall written as in Eq.\ \pref{chi3gen}, where the mixed bubble containing a diamagnetic vertex and a Higgs vertex correspond to $\gamma_{dia}$ for the Higgs. As shown in  Ref.\ \cite{cea_prb16}, $\gamma_{dia}$ is extremely small, suppressing the Higgs contribution to the THG in a clean superconductor. These findings are fully confirmed by the numerical results shown in Fig.\ \ref{jdiav0}. 

In Fig.\ \ref{jdiav0}, as in all the following figures, we show  the modulus of the Fourier transform of the third-order current in the field direction. In panel (a)-(b) we show $|J_{Dia}(\Omega,\theta)|$ as a function of frequency for a fixed polarization $\theta=0,\pi/4$, while in panel (c) we show $|J_{Dia}(\Omega=0,\theta)|$ as a function of the polarization angle $\theta$. The spectrum of $J_{Dia}$ for $\theta=0$, shown in Fig.\ \ref{jdiav0}a,   displays a clear resonance at $\Omega=\Delta$, while the secondary maximum visible  at $\Omega \sim 0.45$  is due to the vHs in the DOS, which is also reported for comparison (grey line). The inclusion of  collective modes barely changes these results. Indeed, as mentioned above, the Higgs mode is weakly coupled to the diamagnetic current, and near half-filling the coupling to charge/phase mode is also negligible in the clean limit\cite{cea_prb16}. This is better seen for  $\theta=\pi/4$, see Fig.\ \ref{jdiav0}b. Here the response is one order of magnitude smaller, and the resonance at $\Omega=\Delta$ is entirely due the Higgs contribution. The same information is encoded in the full angular dependence of $J_{Dia}(\O=0)$ shown in Fig.\ \ref{jdiav0}c. At $\theta=\pi/4$, where only the Higgs mode contributes, the response is negligible with respect to the one at $\theta=0$. By using the decomposition \pref{chidec}, that is allowed for such a Kubo-like response, this is equivalent to state that in the clean case $K_{A1g}\simeq 0$ and the response is fully dominated by the $B_{1g}$ diamagnetic response $J_{Dia}$, with a marked minimum at $\theta=\pi/4$.  
Finally, in Appendix A we compare the present time-dependent results with the one obtained by computing  the THG within the diagrammatic approach, showing the full quantitative agreement among the two techniques in the clean limit.

\section{Disordered system}\label{sec:dis}
\begin{figure*}[htb]
  \includegraphics[width=17cm,clip=true]{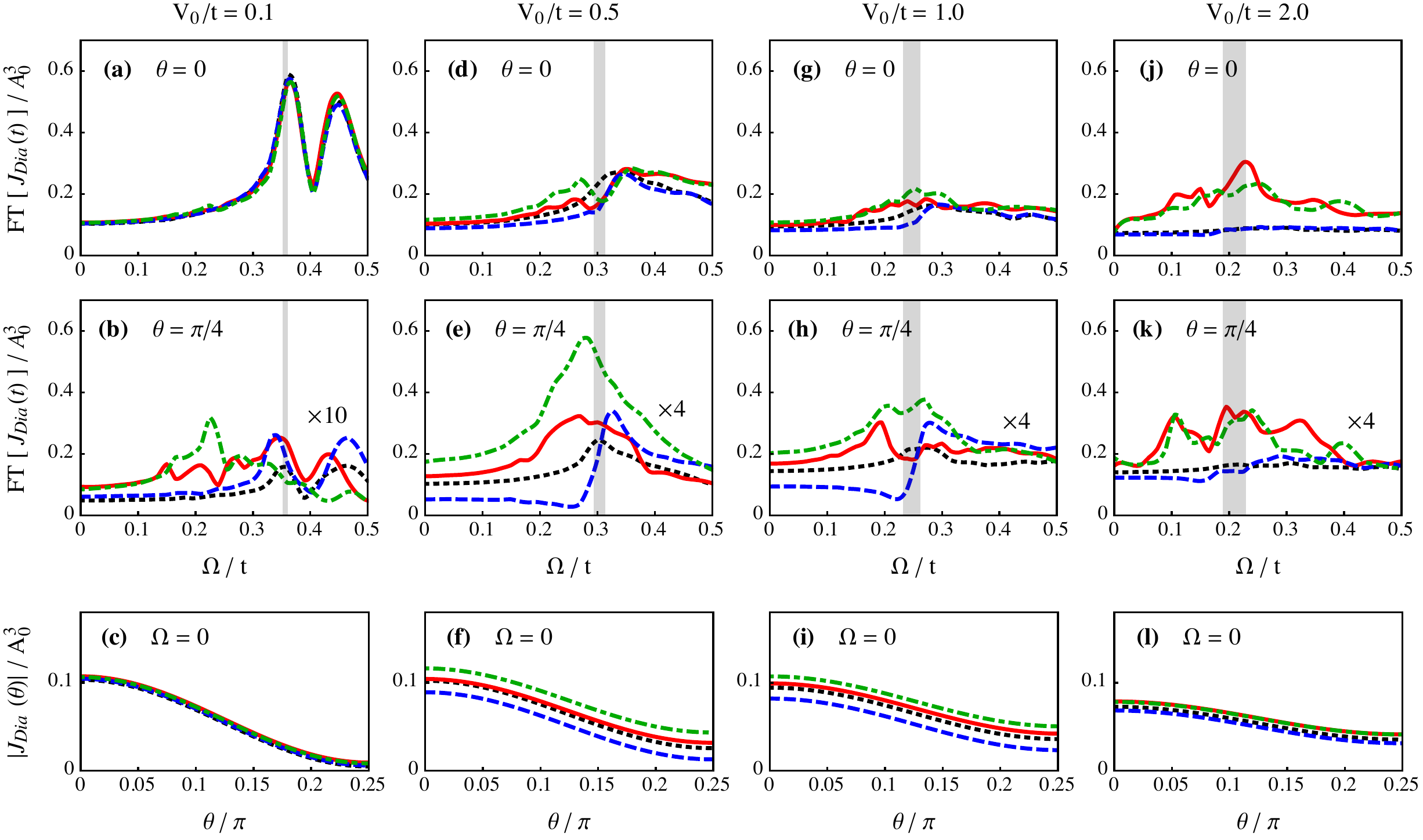}
  \caption{Spectrum of the diamagnetic contribution $J_{Dia}$ to the third-order current at various disorder levels at $\theta=0$ (top row) and $\theta=\pi/4$ (middle row). In each panel vertical bars denote the range of variation of half the spectral gap. As one can see, the results are qualitatively similar to the clean case, shown in Fig.\ \ref{jdiav0}, even though disorder progressively washes out the $\Omega=\Delta$ resonance present at $\theta=0$. Bottom row: polarization dependence of the $\Omega=0$ component of the spectrum. Also in this case the third-order current preserves a minimum at $\theta=\pi/4$, as found without disorder.} 
\label{diaresp}                                                   
\end{figure*}  

\subsubsection{Diamagnetic response}
Fig.\ \ref{diaresp} shows the results for the diamagnetic response $J_{Dia}$ for finite disorder. Focusing first on the $\theta=0$ configuration (top row) one sees that weak disorder leaves the diamagnetic response almost unchanged.  At $V_0/t=0.5$ disorder starts to smear out both the peak at $\Omega=\Delta$ and the one at the vHs.  For large disorder ($V_0/t=2$) the response is dominated by the phase/charge modes, and even though a residual enhancement at $\Omega=\Delta$ is present, a substantial amount of spectral weight  is spread out in the whole frequency range. At $\theta=\pi/4$ the response remains substantially smaller at all disorder levels, as also evidenced by the polarization dependence of the $\Omega=0$ component, shown in the bottom row of Fig.\ \ref{diaresp}.  However, as we shall see below the overall response is much smaller than in the paramagnetic sector, which becomes one order of magnitude larger than the  diamagnetic one already at intermediate disorder. 

\begin{figure*}[bth]
  \includegraphics[width=17cm,clip=true]{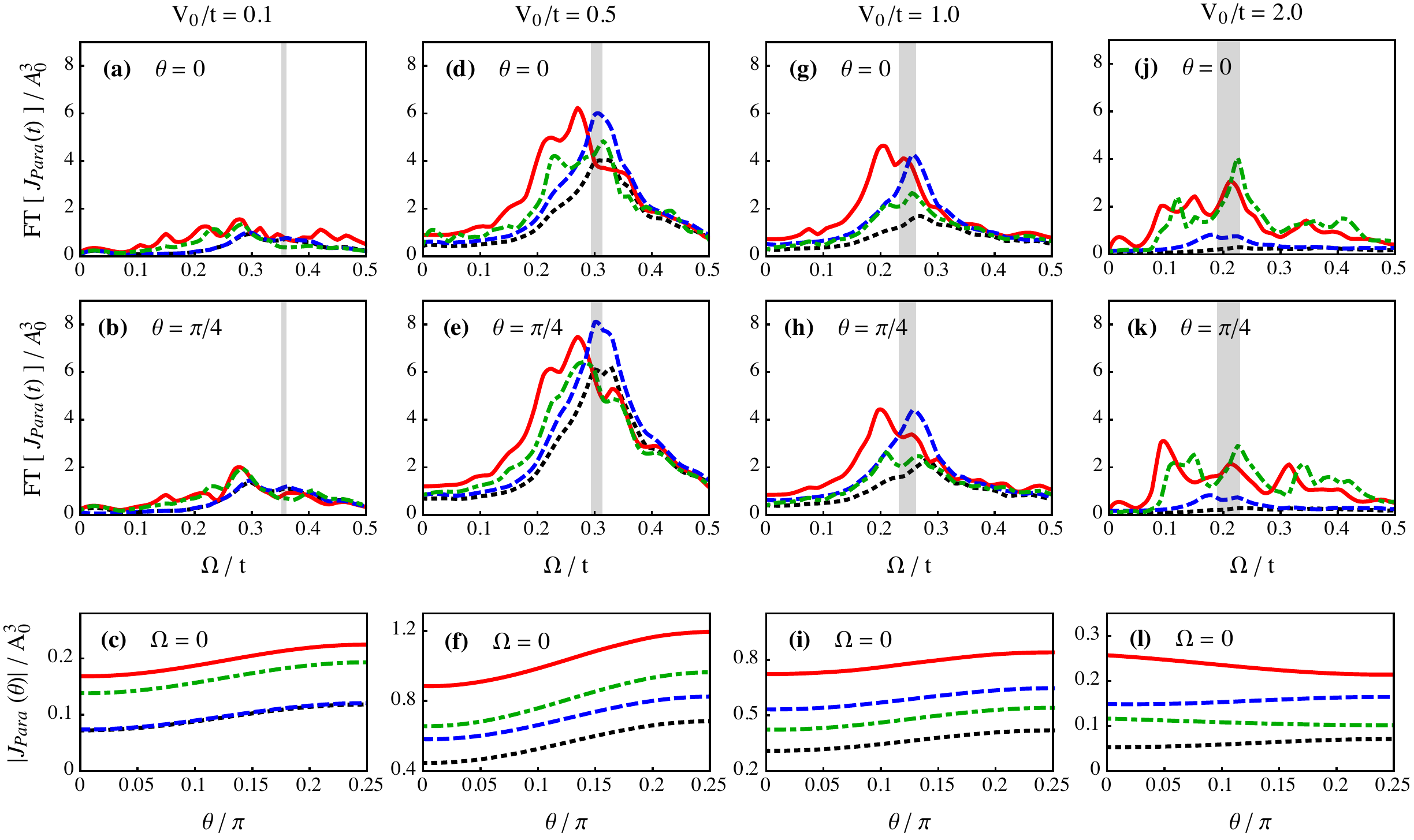}
  \caption{ Spectrum of the paramagnetic  contribution $J_{Para}$ to the third-order current at various disorder levels at $\theta=0$ (top row) and $\theta=\pi/4$ (middle row). In each panel vertical bars denote the range of variation of half the spectral gap. By comparing the scale of the $y$ axis with the one of Fig.\ \ref{diaresp} one sees that already at low disorder $V_0/t=0.5$ the paramagnetic response around $\Omega=\Delta$ is one order of magnitude larger than the diamagnetic one. At all disorder levels, the collective-mode fluctuations give a substantial contribution, with a non-monotomic dependence of the relative intensities as a function of disorder. Bottom row: polarization dependence of the $\Omega=0$ component of the spectrum. In contrast to the diamagnetic case, here the response is maximum at $\theta=\pi/4$, except for the largest disorder level where phase/charge modes tend to restore a very shallow minimum at $\theta=\pi/4$.   }  
\label{figpara}                                                   
\end{figure*}

\subsubsection{Paramagnetic response}
As outlined in Sec.\ \ref{3rd:def} there are two paramagnetic-current contributions to the nonlinear response, $J_{Para}$ and $\tilde J_{Para}$.   As we checked, at all disorder levels $\tilde J_{Para}$ is largely subdominant, as shown explicitly in Appendix \ref{appcurr}. 
The results for $J_{Para}$ as a function of the disorder strength and the polarization are shown in Fig.\ \ref{figpara}. As one can see, already at the lowest disorder level $V_0/t=0.1$ the paramagnetic non-linear response is comparable to the one due to the diamagnetic current shown in Fig.\ \ref{diaresp}, in agreement with previous theoretical work \cite{shimano_prb19,silaev_prb19,tsuji_cm20}. 
As disorder increases further the paramagnetic response largely overcomes the diamagnetic one, with a substantial contribution coming from the collective modes. For intermediate disorder levels $V_0/t=0.5$ and $V_0/t=1$ the inclusion of amplitude fluctuations gives quantitatively the largest effect, even though a direct comparison between the blue (BCS+Higgs) and the red (full response) lines shows that the inclusion of charge/phase fluctuations is not simply additive, and it generally tends to move the maximum towards smaller frequencies. This result can be ascribed to the fact that disorder has also non-trivial effects on the mixing between the Higgs and the phase/charge sectors\cite{cea_prb14,cea_prl15,seibold_prb17,trivedi_prb20}, questioning an estimate of the THG with disorder made without including all the fluctuation channels. Finally, at the strongest disorder level $V_0/t=2$ the charge/phase fluctuations dominate the response in a broad range of frequencies around $\Delta$, in analogy with what already observed for the diamagnetic response. 
By comparing the two configurations at $\theta=0$ and $\theta=\pi/4$ one sees that at low and intermediate disorder the response increases in the diagonal direction, with a progressive softening of the anisotropy, and eventually a reverse of it, as disorder increases. This trend is confirmed by the polarization dependence of the $\Omega=0$ response shown in the bottom row of Fig.\ \ref{figpara}. In particular, when charge/phase modes dominate, the $\O=0$ response is slightly larger at $\theta=0$ than at $\theta=\pi/4$, in contrast with the result found for smaller disorder. 

\section{Discussion and Conclusions}\label{sec:conc}

\begin{figure}[h]
\includegraphics[width=8.cm,clip=true]{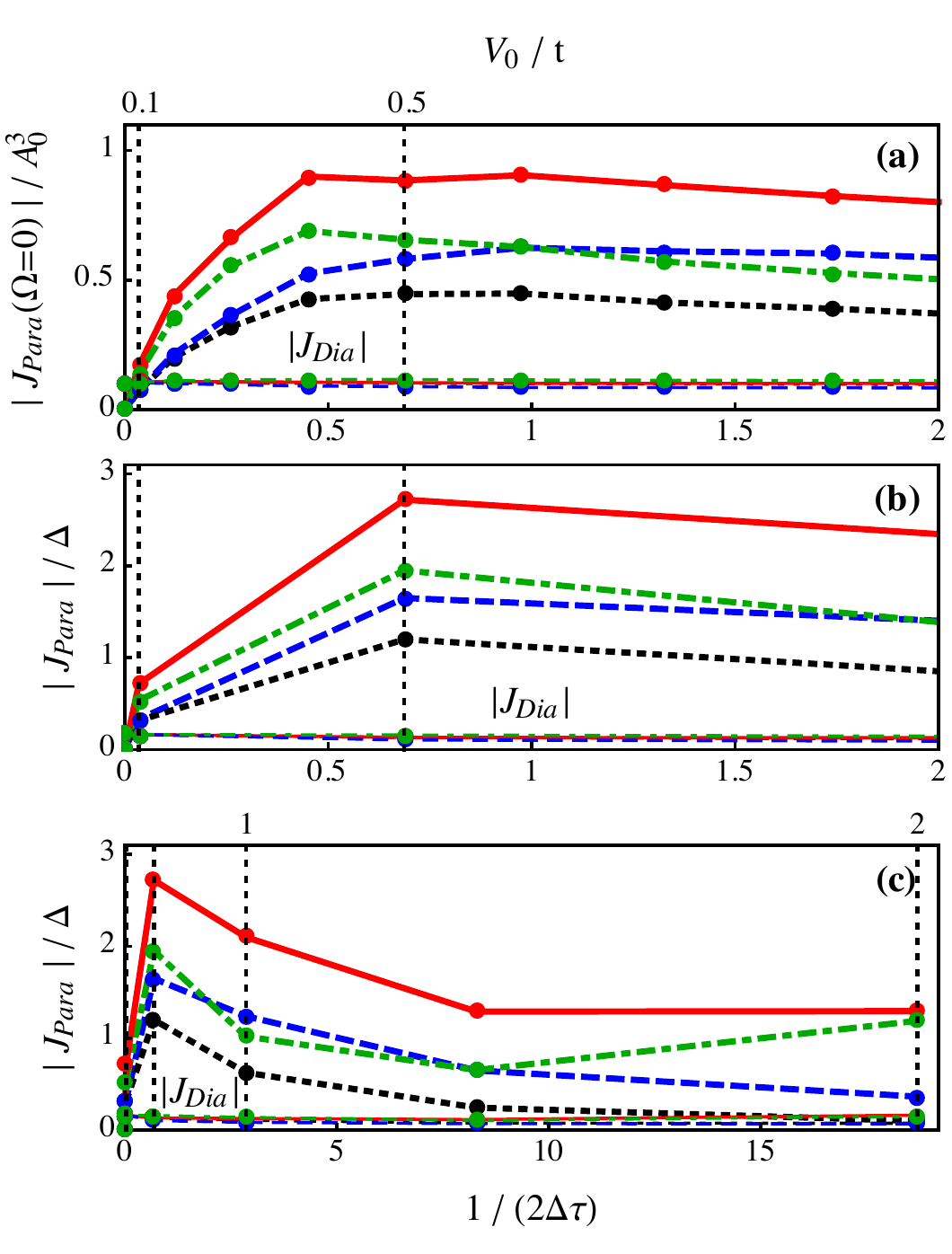}
  \caption{Summary of the results of $J_{Para}$ as a function of disorder, parametrized by $1/(2\Delta\tau)$, with $\tau$ determined from the optical conductivity as outlined in Appendix \ref{appTau}. The correspondence between $V_0/t$ and $1/(2\Delta\tau)$ is reported in Table \ref{tabtau}. Vertical bars denote the values of $V_0/t$ shown in Fig.s \ref{diaresp}-\ref{figpara}, that are also reported on the upper $x$ axis. (a) $\Omega=0$ component of  $J_{Para}$. Here we also report for comparison the $\Omega=0$ component of $J_{Dia}$.  (b) Integrated spectral weight of $J_{Para}(\Omega)$ between $\Omega=0$ and half of the average spectral gap (i.e.\ the grey vertical bars in Fig.s \ref{diaresp}-\ref{figpara}). (c) Same as (b) on a larger disorder range. }
\label{figsumm}                                                   
\end{figure}

To directly compare the various results as a function of disorder we summarize in Fig.\ \ref{figsumm}a the value of the third-order current (for $\theta=0$) at $\Omega=0$, as roughly representative of the strength of the various contributions. To make easier the connection with previous theoretical work and with the experiments, we show the results as a function of $1/(2\Delta \tau)$, where for each value of $V_0$ the corresponding $\tau$ is our estimate of the transport scattering rate based on a fit on the Drude formula \pref{drude}  for the normal-state optical conductivity, and its Mattis-Bardeen counterpart in the SC state, see   Appendix \ref{appTau} for further details. Since previous theoretical work already provided some analytical estimates of the $\Omega=0$ BCS and BCS+Higgs paramagnetic contribution, it is worth comparing such estimates with the present results.
In particular, in Ref.\ \cite{silaev_prb19} the BCS paramagnetic contribution $J_{Para}(\Omega=0)$ has been estimated to scale linearly with $1/(2\Delta\tau)$ at low disorder, while Ref.\ \cite{shimano_prb19,shimano_review19} suggested that it increases initially as $1/(2\Delta \tau)^2$. The results reported in  Fig.\ \ref{figsumm}a are compatible with a linear increase of $J_{Para}(\O=0)$ in the regime $1/(2\Delta\tau)\lesssim 1$, taking into account that we keep fixed $U$ so $\Delta$ also varies slightly with disorder. For what concerns the slope, Ref.\ \cite{silaev_prb19} provides an estimate for a parabolic band dispersion $J_{Para}\simeq 0.06  (v_F^4 N_F/\Delta^2) 1/(2\Delta \tau)$. By using $v_F^4 N_F\simeq t^3$ and the order-of-magnitude $\Delta\simeq 0.3 t$ for the SC order parameter, we obtain $J_{Para}\simeq 0.8/(2\Delta\tau)$, that is roughly in agreement with our numerical results for the BCS contribution. 

\begin{figure}[h]
\includegraphics[width=8cm,clip=true]{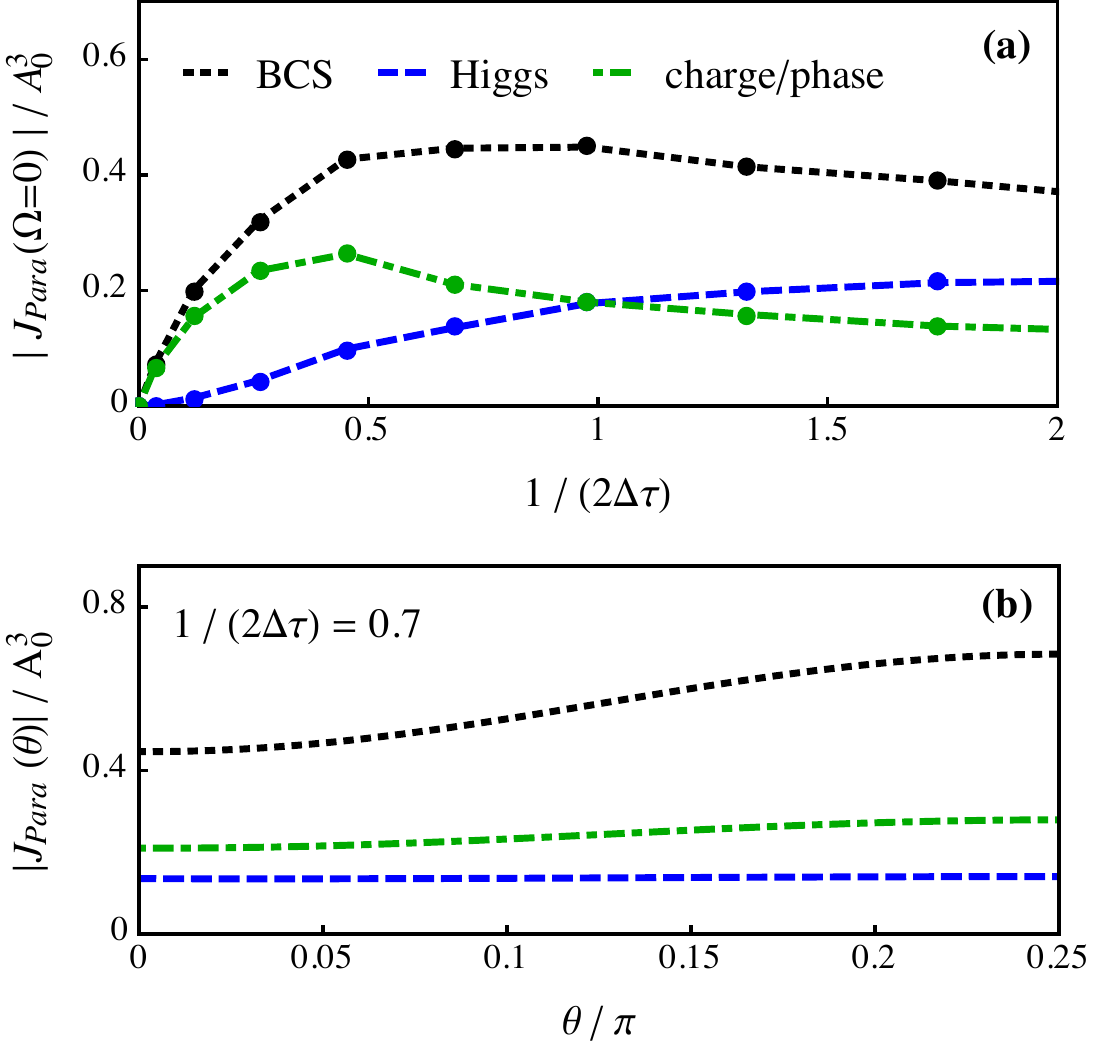}
\caption{(a) Comparison between the BCS response and the collective-modes only response at $\Omega=0$ as a function of disorder. (b) Polarization dependence of the separate contributions at a given disorder level. As one can see, the largest polarization dependence comes from the BCS part, while the collective modes give a rather isotropic contribution. }
\label{figsub}                                                   
\end{figure}  

 Adding the collective modes has a sizeable quantitative effect on the $\Omega=0$ response, with a pronounced contribution of charge/phase fluctuations at very small disorder. By focusing only on the Higgs fluctuations, its contribution starts to be relevant  around $1/(2\Delta \tau)\sim 0.5$, as already observed in Ref.\ \cite{silaev_prb19}, and it becomes comparable to the  phase/charge one as $1/(2\Delta \tau)\gtrsim 1$. These trends are confirmed by analyzing the behavior of the total spectral weight reported in Fig.\ \ref{figsumm}b. Here we show the spectral weight of the third-order current for the configuration $\theta=0$, integrated between $\Omega=0$ and $\Delta$. In the strong disorder regime shown in Fig.\ \ref{figsumm}c the BCS response is largely subdominant and the collective modes dominate the THG. While the effect in the Higgs channel has been already noticed before\cite{silaev_prb19,shimano_prb19,tsuji_cm20}, the contribution of the phase/charge modes has been overlooked so far, even though it clearly dominates the response at very large disorder, which is the one relevant for films of conventional NbN superconductors. On this respect, a detailed analysis including also phase/charge modes for the specific band structure of NbN would be required to finally assess the origin of the THG reported experimentally\cite{shimano_science14,shimano_prb17} for this system.

\begin{figure}[h]
\includegraphics[width=8cm,clip=true]{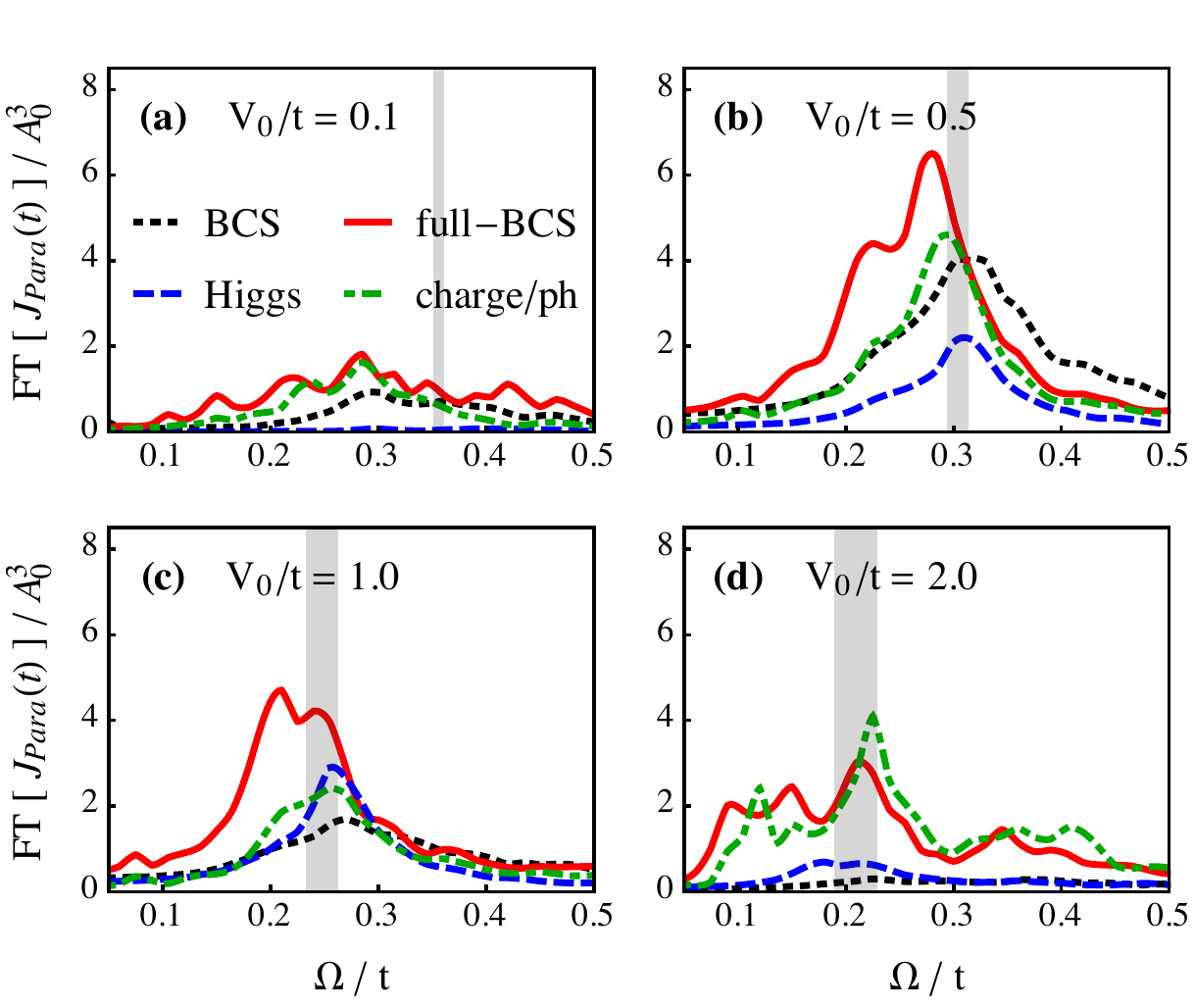}
\caption{(a) Comparison between the BCS response and the collective-modes only response at finite frequency as a function of disorder. }
\label{figsubo}                                                   
\end{figure}  
 
 It is worth noting that here we always report the collective-mode contribution along with the BCS part. Instead in previous work\cite{silaev_prb19,shimano_prb19,shimano_review19,tsuji_cm20} the "Higgs" contribution refers to the resummation in the amplitude channel only, i.e.\ what corresponds to the difference between BCS+Higgs and BCS in our notation. For the $\Omega=0$ response, which is real, this corresponds to just subtracting the values already reported in the previous figures. The result, shown in Fig.\ \ref{figsub}a shows that the THG due only to charge/phase or Higgs fluctuations is still lower than the BCS one up to intermediate disorder levels, in agreement with previous work\cite{silaev_prb19,tsuji_cm20}. At finite frequency the subtraction procedure appears less meaningful, since the $j^{3rd}(\Omega)$ is in general a complex quantity, so the Fourier spectrum of the total response is not necessarily the sum of the Fourier spectra of the separate components. This is evidenced by the Fourier spectra of the separate contributions to $J_{Para}$ shown in Fig.\ \ref{figsubo}, at the same disorder levels  shown in Fig.\ \ref{figpara}.  For what concerns the Higgs-only contribution, it has similar spectral components of the BCS one, and it overcomes the latter at strong disorder $V_0/t\ge 1$ (i.e.\ $1/(2\Delta\tau)\geq 3$.) On the other hand, the charge/phase mode has a substantial contribution from spectral components below the gap, in analogy with what found within the context of the optical conductivity\cite{cea_prb14,seibold_prb17,pracht_prb17}. As a consequence, when charge/phase modes are included the subtraction procedure highlights an enhancement of the THG towards a value slightly smaller than $\Delta$.  It is worth noting that in the present model we did not include explicitly long-range Coulomb forces, so the phase mode preserves a sound-like dispersion in the homogeneous case. As it is well known, Coulomb interactions modify the spectrum of the phase/charge mode, pushing it towards the plasma frequency. Nonetheless, the Goldstone theorem is still preserved, since in the long wavelength limit $\bq=0$ phase fluctuations are still costless\cite{nagaosa}, and in addition in a strictly two-dimensional systems the plasmon mode will still start from zero with a $\sqrt{q}$ dependence.  On the other hand, the effect of Coulomb interactions on phase fluctuations at large q could be expected to be small. 
  In the presence of disorder one mixes to some extent the fluctuations at different momenta, so it is not evident a-priori how the present results will be quantitatively and qualitatively modified in the presence of long-range forces. This is an interesting issue that will be deserved for future investigation.

 Finally, we would like to comment on the results for the polarization dependence of the signal in connection to recent observations in unconventional cuprate superconductors\cite{shimano_prl18,kaiser_natcomm20,shimano_prb20}. As mentioned at the beginning, our tight-binding model on the square lattice is a first approximation for the band structure of cuprates. Even though we did not consider explicitly a $d$-wave symmetry for the SC order parameter, as appropriate for cuprates, we expect that it will not affect significantly our results for what concerns the effect of disorder on the polarization dependence of the THG. Indeed, as shown by previous work\cite{cea_prb18,shimano_prb17,tsuji_cm20}, this is mainly controlled by the band-structure effect, while the gap symmetry influences the spectrum of the Higgs mode itself, making it possible excitations also below twice the gap maximum\cite{manske_natcomm20}. As detailed in Sec.\ \ref{sec:THG}, even small disorder immediately triggers a large paramagnetic response with a polarization dependence opposite to the one found in the clean system. 
 The cuprate films investigated experimentally so far\cite{shimano_prl18,kaiser_natcomm20,shimano_prb20} are expected to be in the relatively clean limit $1/(2\Delta\tau)\sim0.85$, as evidenced by the linear decrease of the superfluid density at low temperature and by the small scattering rate $\tau\simeq 0.2$ ps obtained by fitting the THz optical conductivity\cite{shimano_prb20}. In this regime, our results show that  the non-linear optical response is still dominated by the BCS part of the $J_{para}$ current, even though all collective modes contribute to enhance the response. However, the experimentally-observed polarization dependence is still puzzling. Indeed, recent results in various families of cuprates, obtained either by directly measuring the THG\cite{kaiser_natcomm20} or the pump-probe signal\cite{shimano_prl18,shimano_prb20} showed a marked isotropic contribution, with a residual angular dependence that,  especially in overdoped Bi2212 samples\cite{shimano_prl18}, gives a  minimum of the response at $\theta=\pi/4$. 
 The estimated disorder level corresponds roughly to $V_0/t=0.5$ in Fig.\ \ref{figpara}. In this case one would expect a signal that is weakly angle dependent, but with a maximum of the response at $\theta=\pi/4$, see Fig.\ \ref{figpara}f.  
These findings suggest that additional effects beyond disorder, specific to cuprates, could be relevant. One possibility is the THG due to collective-mode contributions beyond RPA, as e.g.\ the one due to two-plasmon processes, recently discussed in Ref.\ \cite{gabriele_cm20}. Indeed, as shown in Ref.\ \cite{gabriele_cm20}, these processes have a sizeable $B_{1g}$ component that can compensate the modulation due to paramagnetic processes shown in Fig.\ \ref{figpara}. A quantitative estimate of the relative strength of the various processes would be certainly interesting, even though  it represents a non-trivial extension of our approach that goes well beyond the scope of the present manuscript. 

In summary, we have investigated the third harmonic response of
disordered superconductors within a real-space time-dependent BdG approach.
This method allows us to treat disorder exactly, and to add the effects of all collective-mode fluctuations in the superconducting state, i.e.\ the Higgs, phase and charge modes. In addition, by considering explicitly a lattice band structure we have been able to highlight the effects of disorder on the polarization dependence of the third-order current. For weak and  intermediate disorder we find that the BCS response still provides the main  contribution to the THG.  However, in agreement with previous work\cite{silaev_prb19,shimano_prb19}, we found that it is driven by paramagnetic instead of diamagnetic contributions, which dominate instead the clean limit\cite{cea_prb16}. A first consequence is that the polarization dependence of the signal changes drastically with respect to the homogeneous case. For the band structure considered in the present manuscript, which is in first approximation appropriate for cuprate superconductors, the BCS response changes from having a marked maximum at $\theta=0$ in the homogeneous case to having an almost isotropic behavior with a small relative maximum at $\theta=\pi/4$ in the disordered case. A second crucial observation is that paramagnetic electronic processes mediate a contribution from {\it all} collective modes, not only the Higgs one studied so far. In our model the effect of phase/charge modes is always as large as the Higgs one. In the regime $1/(2\Delta\tau)\geq 1$ the full response including all fluctuations remains peaked around $\Delta$, i.e.\ half of the spectral gap, while in the strong disorder limit $1/(2\Delta\tau)\gg1$ charge/phase modes tend to contribute to a wide spectral range below $\Delta$. The finite contribution of charge/phase modes to the non-linear current is reminiscent of the same result already established for the linear current\cite{cea_prb14,seibold_prb17,pracht_prb17}, that has been invoked to explain the anomalous sub-gap absorption in strongly disordered films of conventional superconductors\cite{armitage_prb07,driessen_prl12,dressel_natphys15,armitage_prb16,scheffler_prb16}. In the case of cuprate superconductors, which belong to the low-disorder limit,  our results suggest that the paramagnetic BCS response gives the largest contributiont, even though all collective modes cooperate to the THG enhancement below $T_c$. Interestingly, the polarization dependence of the signal is still partly in disagreement with the available experimental data, which also show a non-trivial dependence on the cuprate family and on the doping level\cite{shimano_prl18,kaiser_natcomm20,shimano_prb20}. Whether this discrepancy can be solved by a more realistic band structure, and a doping-dependent effective disorder, or by including fluctuation effect beyond RPA\cite{gabriele_cm20}, is an open question that certainly deserves further theoretical and experimental investigation.

%
%
\acknowledgements
This work has been supported by the Sapienza University
via Ateneo 2019 RM11916B56802AFE, by the Italian MIUR project PRIN 2017 No.
2017Z8TS5B, and by Regione Lazio (L. R. 13/08) under
project SIMAP. G.S. acknowledges financial support from the Deutsche Forschungsgemeinschaft under SE 806/19-1.

\appendix

\section{Homogenous case: comparison with the Anderson pseudospin formalism and the effective-action approach}\label{appHom}

Let us first show the equivalence between our approach and the Anderson pseudospin description in the clean case. For a homogeneous system the lattice momentum $\bk$ is a good quantum number and by introducting the Fourier transform of the electron creation and annihilation operators $c^\dagger_\bk,c_\bk$, such that $c_{\bk\sigma}=\sum_i e^{-i\bk\cdot \bR_i}c_{i\sigma}$ one can express also the BdG transformations \pref{cidef} with the usual BCS $u^2_\bk$ and $v^2_\bk$ coherence factors:
\bea
\lb{uk}
u^2_\bk&=&\frac{1}{2}\left( 1+\frac{\xi_\bk}{E_\bk}\right),\\
\lb{vk}
v^2_\bk&=&\frac{1}{2}\left( 1-\frac{\xi_\bk}{E_\bk}\right),
\eea
such that the eigenvalues $\omega_k$ in Eq.\ \pref{eq1}-\pref{eq2} correspond to the BCS energies $E_\bk=\sqrt{\xi_\bk^2+\Delta^2}$, where $\xi_\bk=\epsilon_\bk-\mu$, $\epsilon_\bk=-2t(\cos k_x+\cos k_y)$ is the band structure corresponding to Eq.\ \pref{eq:hub} for nearest-neighbor hopping $t$ only, and the SC order parameter $\Delta$ is given by
\be
\Delta=-|U|\sum_\bk \langle c_{-\bk\down} c_{\bk\up}\rangle= -\sum_\bk u_\bk v_\bk.
\ee
For the translationally invariant case it is convenient to express also the elements of the density matrix $\cal R$ in momentum space, by introducing the variables:
\be
\lb{rk}
{\cal R}(\bk) = \left(
\begin{array}{cc}
\langle c_{k,\uparrow}^\dagger c_{k,\uparrow}\rangle  & \langle c_{k,\uparrow}^\dagger c_{-k,\downarrow}^\dagger\rangle \\
\langle c_{-k,\downarrow}c_{k,\uparrow}\rangle &
\langle c_{-k,\downarrow}c_{-k,\downarrow}^\dagger\rangle
\end{array}\right)\,.
\ee
The corresponding equations of motion have then the same structure of Eq.\ \pref{eq:mot}, as given by:
\begin{equation}
\lb{drk}
 i \frac{d}{dt}{\cal R}(\bk)=\left\lbrack {\cal R} (\bk), H^{BCS}(\bk) \right\rbrack 
  \end{equation}
  where $H^{BCS}(\bk)=\langle \psi^\dagger_\bk \hat H_\bk \psi_\bk\rangle$ is the average value over the BCS ground state $|{BCS}\rangle$ of the standard BCS Hamiltonian $\hat H_\bk$, written in the basis of the Nambu spinor $\psi^\dagger_\bk=(c^\dagger_{\bk\up}, c_{-\bk\down})$:
\be
\hat H_\bk= 
\left(
  \begin{array}{cc}
    \xi_\bk & \Delta \\
    \Delta^* & -\xi_{-\bk}  \end{array} \right) .
    \ee    
Eq.\ \pref{drk} can be also expressed in a completely equivalent form by introducing the vector
\be
\lb{defb}
\bb_\bk=
\left( 
\begin{array}{c}
\Delta_x \\ \Delta_y \\ \xi_\bk
\end{array} \right)
\ee
where $\Delta=\Delta_x-i\Delta_y$, and by using suitable combinations of the elements of ${\cal R}(\bk)$ by defining a vector $\bS_\bk$ as: 
\be
\lb{defs}
\bS_\bk=\frac{1}{2} \langle \psi^\dagger_\bk \vec \sigma \psi_\bk \rangle
\ee
with $\vec\sigma=(\sigma_x,\sigma_y,\sigma_z)$ denoting the Pauli matrices. By using the dynamical variables \pref{defs} the $H^{BCS}(\bk)$ can be expressed as:
\be
H^{BCS}(\bk)=2\bb_\bk \cdot \bS_\bk
\ee
and the equations of motion \pref{drk} assume the compact form
\be
\lb{prec}
\frac{d}{dt}\bS_\bk=2\bb_\bk \times \bS_\bk 
\ee
that reminds the precession of a pseudospin $\bS_\bk$ around an external fictitious magnetic field $\bb_\bk$. This is the so-called Anderson pseudospin formalism employed in previous work\cite{shimano_science14,aoki_prb15} to study the THG for clean superconductors. 
Once that a homogeneous gauge field $\bA$ is introduced by minimal-coupling substitution to the fermionic operator, the $z$ component of the $\bb_\bk$ vector in Eq.\ \pref{defb} gets replaced by $(\xi_{\bk-e\bA}+\xi_{\bk+e\bA})/2$, and the solution of the equation of motion \pref{prec} at finite $\bA$ can be used to compute the current. In particular, for a field $A_x(t)$ in the $x$ direction the total current up to third order in $A_x$ is defined by Eq.\ \pref{eq:jfexp} above,  that we report here for convenience:
\begin{eqnarray}
  \lb{eqjtot}
  j(t)=(1-\frac{1}{2}A_x^2)j_{para}(A)+(A_x-\frac{1}{6}A_x^3)j_{dia}(A) \quad 
  \end{eqnarray}
For the translationally invariant case the total paramagnetic and diamagnetic current can be expressed using either the components of ${\cal R}(\bk)$ or the pseudospin components $\bS_\bk$, i.e.:
\bea
j_{para}(t)&=&-2t\sum_\bk \sin k_x ({\cal R}_{11}(\bk)+{\cal R}_{22}(\bk))=\nn\\
\lb{jparak}
&=&-4t\sum_\bk \sin k_x S_0(t)\\
j_{dia}(t)&=&-2t\sum_\bk \cos k_x ({\cal R}_{11}(\bk)-{\cal R}_{22}(\bk))=\nn\\
\lb{jdiak}
&=&-4t\sum_\bk \cos k_x S_z(t)
\eea
where we introduced a fourth component $S_0\equiv \frac{1}{2}\langle \psi^\dagger_\bk \psi_\bk\rangle$ to describe the paramagnetic current. In Eq.\ \pref{jparak}-\pref{jdiak} the time dependence of the pseudospin operators is induced by the applied gauge field $A_x(t)$. Indeed, when $A_x=0$ one can easily see that choosing e.g.\ a real order parameter $\Delta$ the pseudospin vector $\bS_\bk(0)=-(\Delta,0,\xi_\bk)/2E_\bk$ and the fictitious field $\bb_\bk=(\Delta, 0, \xi_\bk)$ are parallel, so that $\bb_\bk\times \bS_\bk=0$ and from Eq.\ \pref{prec} one obtains $\bS_\bk(t)=\bS_\bk(0)$. As a consequence, only the diamagnetic current \pref{jdiak} is different from zero and constant, with 
$ j_{dia} =2t\sum_\bk (\cos k_x) \xi_\bk/2E_\bk$. In this case the linear term in $A_x$ of Eq.\ \pref{eqjtot} gives the usual superfluid response, while the third-order current admits only the instantaneous response. When a finite gauge field is present $\bb_\bk(\bA)\times \bS_\bk\neq 0$, and the pseudospin $\bS_\bk(t)$ acquires a finite dynamics. By expanding the $\bb_\bk(\bA)$ fictitious field at quadratic order in $A_x$ one can derive analytically for the homogeneous case the various contributions to the third-order current.
When the frequency of $\bA(t)$ is in resonance with the gap,
  the energy transfer of the external field to the SC system induces a singular
  growth of the third-order current response. We have therefore supplemented
 Eq. (\ref{prec}) with a phenomenological damping
\be
\lb{prec2}
\frac{d}{dt}\bS_\bk=2\bb_\bk \times \bS_\bk -\tilde{\gamma} (\bS_\bk-\bS^{(0)}_\bk)
\ee
which constraints the dynamics and allows us for a better comparison with the
diagrammatic approach, where the analogous divergence can be also regularized
by a phenomenological damping parameter (see below).
Further details on the third-order current derivation within a perturbative psuedospin approach can be found in Ref.\ \cite{shimano_science14,aoki_prb15,cea_prb16}.

A second possible way to derive the third-order current is to follow a diagrammatic approach based on the effective-action formalism, as done in Ref.\ \cite{cea_prb16,cea_prb18,udina_prb19}. In this case one builds up by means of the Hubbard-Stratonovich decoupling an effective action for the collective SC degrees of freedom (i.e.\ amplitude, phase and charge fluctuations). After introducing the gauge field and integrating out the collective modes one can derive an effective action $S^{(4)}(\bA)$ up to fourth order in $\bA$. Diagrammatically, this corresponds to compute the third-order current by including the BCS response and the contribution of all collective modes at RPA level. This procedure is completely equivalent to compute the dynamical response within the Anderson pseudospin model. Indeed, both approaches confirm the results of Sec.\  \ref{THGclean}, i.e.\ that for the clean system only the  diamagnetic response is present, and Higgs fluctuations give a negligible contribution.  The initial suggestion of Ref.\ \cite{shimano_science14} of a predominant Higgs contribution was indeed motivated by an incorrect evaluation of the total current within the Anderson pseudospin approach,  as shown in details in the Supplementary paper of Ref.\ \cite{cea_prb16}.
\begin{figure}[h]
\includegraphics[width=8cm,clip=true]{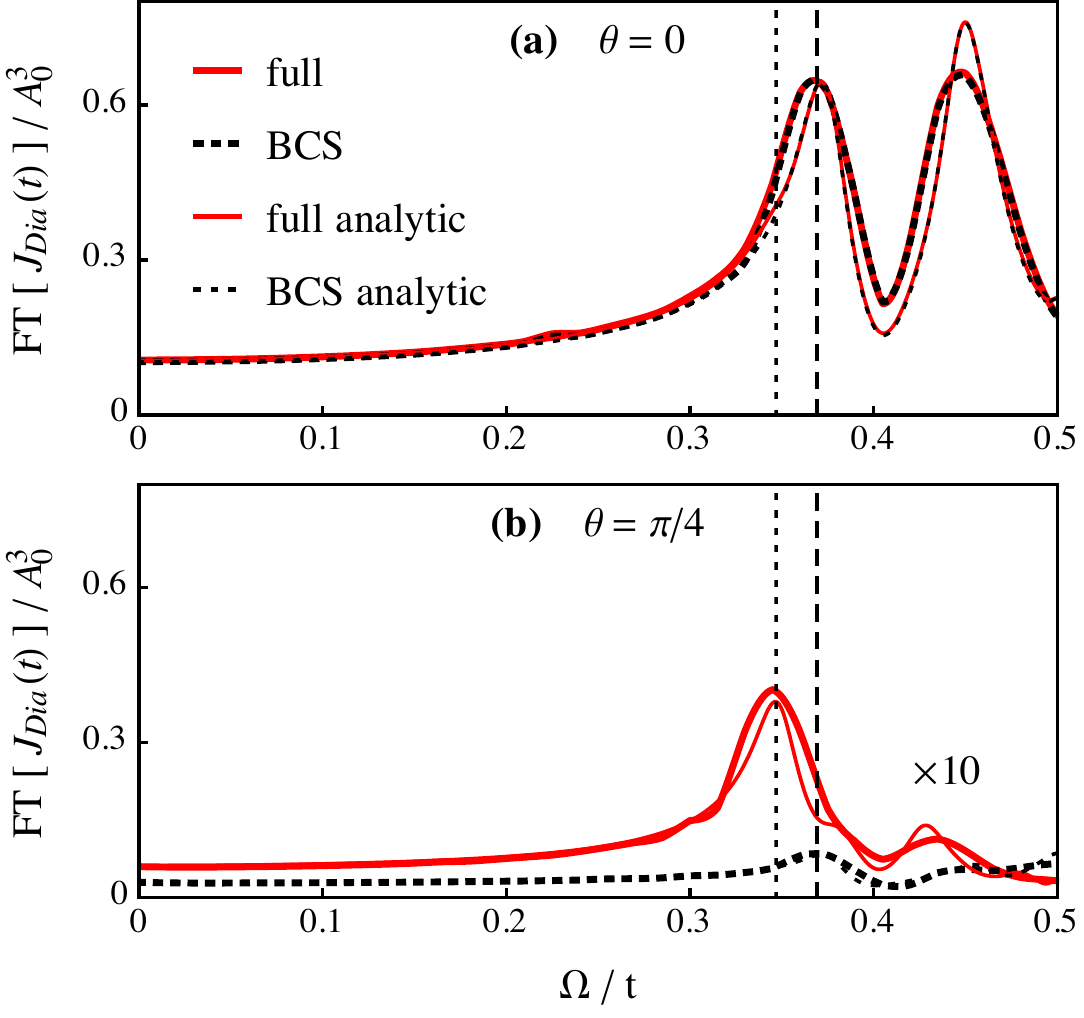}
\caption{Comparison between the numerical results obtained by the numerical solution of the density-matrix equations of motion, shown also in Fig.\ \ref{jdiav0}, and the analytical expressions \pref{jBCS}-\pref{jCM} for the same system size, labelled as BCS analytic and full analytic. As one can see, there is an excellent quantitative agreement for both polarizations.  As in Fig.\ \ref{jdiav0}, the vertical dashed line indicates half of the spectral gap, while the vertical dotted line denotes the SC order parameter. Here we used  $\tilde{\gamma}/\Delta=0.05$ for the
  time-dependent approach and $\gamma/\Delta=0.07$ for the analytical one. }
\label{figcomp}                                                   
\end{figure}  

Once established the full equivalence between the time-dependent approach based on the density matrix, used in the present manuscript, and previous work based either on the Anderson pseudospin description or on the effective-action formalism, we can now compare our numerical results for the clean case with the analytical expressions  derived previously. For the sake of definitiviness, we will use the same notation of Ref.\ \cite{cea_prb16}, where the same band structure was investigated. For a gauge field $\bA$ applied at a generic angle $\theta$ with respect to the $x$ axis the Fourier transform of the third-order current can  be written as:
\bea
j^{(3)}_{BCS}(\theta,\Omega)/A_0^3&=&\frac{1}{2}\Big[ \chi^{BCS}_{xx}(2\Omega)(\cos^4\theta+\sin^4\theta)+\nn\\
\lb{jBCS}
&+&2\chi^{BCS}_{xy}(2\Omega)\sin^2\theta\cos^2\theta \Big]
\eea
where we defined
\be
\lb{chiBCS}
\chi^{BCS}_{ij}(\Omega)=-\frac{\Delta^2}{N}\sum_{\bf{k}} \partial_i^2\epsilon_{\bf{k}} \partial_j^2\epsilon_{\bf{k}} F_{\bf{k}}(\Omega)
\ee
and we put
\bea
\lb{deffk}
F_{\bf{k}}(\Omega)&=&\frac{1}{E_{\bf{k}}\left[4E_{\bf{k}}^2-(\Omega+i\gamma)^2\right]} 
\eea 
From Eq.\ \pref{chiBCS}-\pref{deffk} one clearly sees the divergence of the BCS response function at $\Omega=2\Delta$, that corresponds to the enhancement of the THG  at $\Omega=\Delta$, since the diamagnetic current \pref{jBCS} is computed at twice the field frequency. As mentioned above, we added in Eq.\ \pref{deffk}  a finite damping $\gamma$ to regularize such a divergence. The contribution of the amplitude, phase and charge collective modes (CM) is independent on the polarization angle $\theta$ and it can be expressed as:
\bea
j^{(3)}_{CM}(\theta,\Omega)/A_0^3&=& - \frac{\chi_{A^2\rho}^2}{2\chi_{\rho\rho}}-
\frac{\left( \chi_{A^2\Delta} - \chi_{A^2\rho} \chi_{\rho \Delta}/\chi_{\rho\rho}\right)^2}{2X_{\Delta\Delta}-2\chi_{\rho\Delta}^2/\chi_{\rho\rho}}\nn\\
\lb{jCM}
\eea
where we defined the following susceptibilities
\bea
\chi_{\rho\rho}(\Omega)&=& -\frac{4\Delta^2}{N}\sum_{\bf{k}}F_{\bf{k}}(\Omega) ,\nn\\
\chi_{\rho\Delta}(\Omega)&=& -\frac{4\Delta}{N}\sum_{\bf{k}}\xi_{\bf{k}} F_{\bf{k}}(\Omega) , \nn\\
\chi_{A^2\rho}(\Omega)&=& -\frac{2\Delta^2}{N}\sum_{\bf{k}}\partial_i^2\epsilon_{\bf{k}}F_{\bf{k}}(\Omega) ,\nn\\
\chi_{A^2\Delta}(\Omega)&=& -\frac{2\Delta}{N}\sum_{\bf{k}}\partial_i^2\epsilon_{\bf{k}} \xi_{\bf{k}} F_{\bf{k}}(\Omega).
\eea
Here $\chi_{\rho\rho}$ is the BCS charge susceptibility, $\chi_{\rho\Delta}$ is the coupling between the phase/charge and amplitude fluctuations, and  $\chi_{A^2\Delta}$ and $\chi_{A^2\rho}$ denote the coupling between the gauge field and the Higgs or phase/charge sector, respectively. Finally, the amplitude fluctuations are described by the inverse of $X_{\Delta\Delta}$, where 
\be
X_{\Delta\Delta}(\Omega) = \frac{1}{N} \left[(2\Delta)^2- \Omega^2 \right] \sum_{\bf{k}} F_{\bf{k}}(\Omega)
\ee
One can easily see that at half-filling $\chi_{\rho\Delta}=\chi_{A^2\rho}=0$ by symmetry, so that amplitude and phase sectors are decoupled, and only the Higgs fluctuations contribute to the THG. Eq.\ \pref{jCM} can then be simplified as:
\be
j^{(3)}_{Higgs}= - \frac{\chi_{A^2\Delta}^2}{2X_{\Delta\Delta}}
\ee
so that the total third-order current is given by $j^{(3)}_{full}=j^{(3)}_{BCS}+j^{(3)}_{Higgs}$. At the doping $n=0.875$ we are considering in the present manuscript  one can still approximate $\chi_{\rho\Delta}\simeq \chi_{A^2\rho}\simeq 0$, and only the BCS+Higgs contribution survives. In Fig.\ \ref{figcomp} we compare the results obtained numerically in the clean case by our density-matrix approach and the analytical expression \pref{jBCS}-\pref{jCM} computed for a system of the same size $N\times N$ of our simulations, with $N=16$. As one can see, the quantitative agreement between the two procedures is excellent. As mentioned in the caption of Fig.\ \ref{jdiav0}, for small systems as the one we considered there can be a difference in the clean case between the SC order parameter and half of the spectral gap, identified by the maximum of the DOS. This is due to the fact that for finite systems the chemical potential $\mu$, computed self-consistently for the same lattice, does not necessarily coincide with an energy state $\varepsilon_k$. As a consequence the spectral gap, which is given by definition by the minimum value of $2E_k=2\sqrt{\Delta^2+(\varepsilon_k-\mu)^2}$, does not necessarily coincides with $2\Delta$. Such a difference is also present in the analytical formula computed at small $N$, and we checked that it disappears as expected at large $N$. In the disordered case the spectral gap becomes intrinsically inhomogeneous, washing out the relevance of such finite-size effects.

\section{Additional details on the numerical procedure}\label{appNum}

To obtain the full current in Eq.\ \pref{eq:j3rd} we need to compute the various orders $j_{dia/para}^{(m)}$ of the diamagnetic/paramagnetic current. This decomposition can be obtained by evaluating $j_{para/dia}$ for different
numerical values of $A_0$. In fact, computing Eq. (\ref{eq:jpexp}) for a vector
potential $A_0'$ and $A_0''$ one finds
\begin{eqnarray}
  j_{para/dia}^{(1)}(t) &=& \left\lbrack \frac{j_{para/dia}^{full}(t)}{A_0}\right\rbrack
  - A_0 j_{para/dia}^{(2)} \nonumber \\ &-& A_0^2 j_{para/dia}^{(3)} + \dots \label{eq:jpd1}\\
  j_{para/dia}^{(2)}(t) &=& \frac{1}{A_0-A_0'}\left\lbrace\left\lbrack \frac{j_{para/dia}^{full}(t)}{A_0}\right\rbrack -\left\lbrack \frac{j_{para/dia}^{full}(t)}{A_0'}\right\rbrack\right\rbrace \nonumber \\
  &-& (A_0+A_0') j_{para/dia}^{(3)}+ \dots \label{eq:jpd2}\\
  j_{para/dia}^{(3)}(t) &=& \frac{1}{(A_0-A_0')(A_0-A_0'')}\left\lbrack \frac{j_{para/dia}^{full}(t)}{A_0}\right\rbrack \nonumber\\
  &-& \frac{1}{(A_0-A_0')(A_0'-A_0'')}\left\lbrack \frac{j_{para/dia}^{full}(t)}{A_0'}\right\rbrack \nonumber\\
  &+&\frac{1}{(A_0-A_0'')(A_0'-A_0'')}\left\lbrack \frac{j_{para/dia}^{full}(t)}{A_0''}\right\rbrack \nonumber\\
  &+& {\cal{O}}(A_0) j_{para/dia}^{(4)}+ \dots \label{eq:jpd3}
\end{eqnarray}
and the current in the square brackets is evaluated with the $A_0$ as denoted
in the corresponding denominator.
Thus for small enough values of $A_0$, $A_0'$, $A_0'', \dots$ one can extract the
various orders in this way and check whether the extracted Fourier components
are independent of the specific choice of the vector potential amplitudes. 

By using the above procedure we computed  the various contributions to the third-order current in Eq. (\ref{eq:j3rd}) at various disorder strength for a $16\times 16$ lattice, by fixing the particle density $n=0.875$ and the local attraction $-|U|=-2t$.  The time-dependent current is calculated
over $10$ periods of the applied vector potential $A_x(t)=A_0\sin(\Omega t)$
and three values of $A_0=10^{-3}$, $10^{-4}$, $10^{-5}$ which then
are used to calculate $j_{para/dia}^{(1,2,3)}$ from Eqs. \pref{eq:jpd1}-\pref{eq:jpd3}). The results are averaged over $20$ disorder configurations.

\section{Subleading contributions to the third-order current}\label{appcurr}

\begin{figure}[hhh]  
\includegraphics[width=8cm,clip=true]{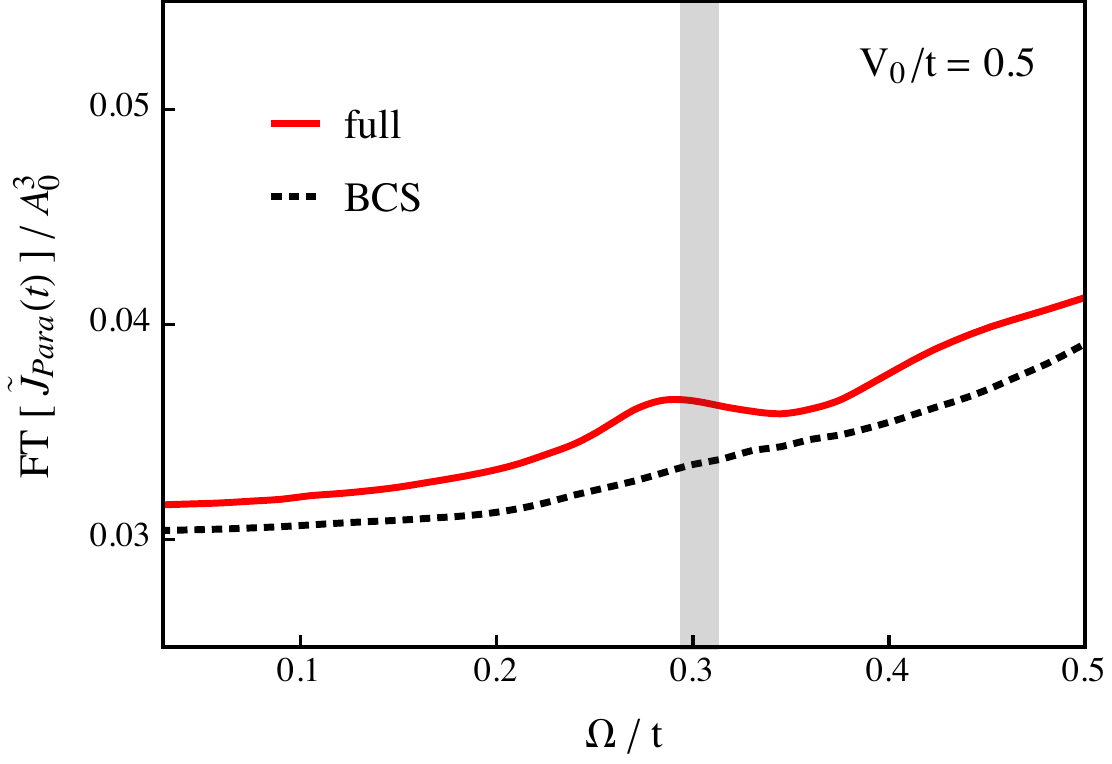}
  \caption{BCS (dashed black line) and full (solid red line) contribution to the third order current $\tilde J_{Para}$. As usual, the BCS result is the one obtained without including any collective-mode fluctuation. } 
\label{figpara1v05}                                                   
\end{figure}

As we mentioned in Sec.\ \ref{sec:THG}, in the presence of disorder the two contributions $\tilde J_{Para}$ and $J_{Inst}$ are subleading with respect to the paramagnetic response $J_{Para}$. For what concerns $\tilde J_{Para}$, it originates from the same Kubo-like current-current correlation function responsible for the finite optical absorption in a disordered superconductor, see second row in Fig.\ \ref{diag}. This term, as emphasized before, is only present in a lattice model. In 
Fig.\ \ref{figpara1v05} we show as an example the results of $\tilde J_{Para}$ for disorder strength $V/t=0.5$. As one can see, this contribution is smaller by almost two orders of magnitude than $J_{para}$ as shown in Fig.\ \ref{figpara}. Similarly to the optical conductivity, we have also checked that $\tilde J_{Para}$ is identical, within the numerical accuracy, to the result obtained directly with the Kubo approach\cite{cea_prb14,seibold_prb17}.

\begin{figure}[hhh]  
\includegraphics[width=8.5cm,clip=true]{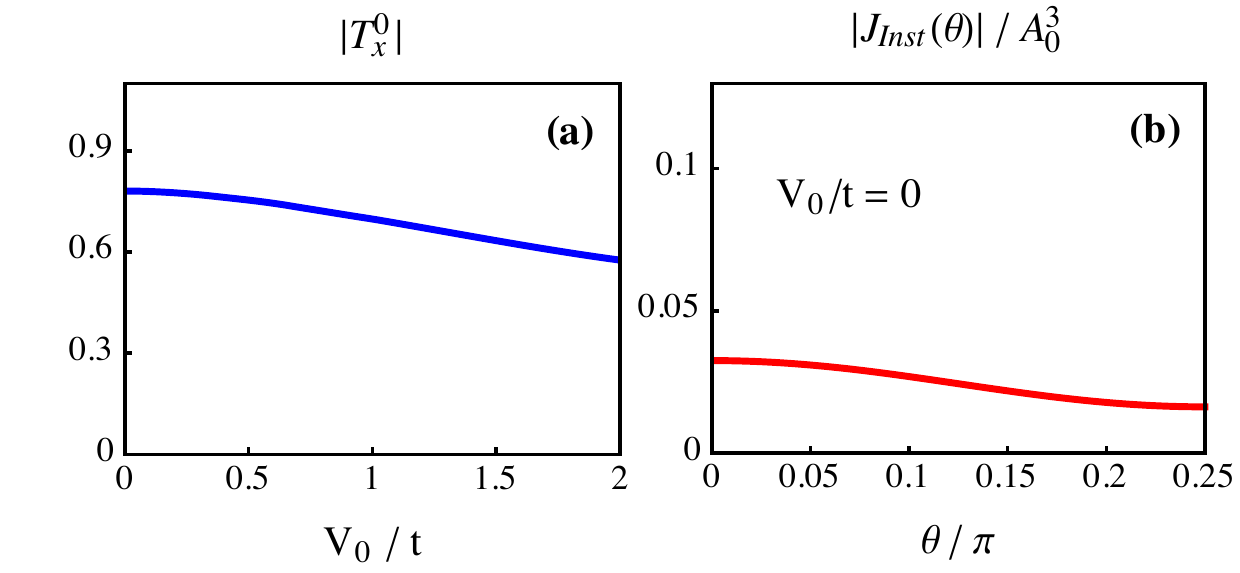}
  \caption{(a) Total kinetic energy along the $x$ direction as a function of the disorder strength. (b) Polarization dependence of the instantaneous response $J_{Inst}$ (\ref{jinstpll}). Here we used the value $T^0_x$ obtained in the clean case. } 
\label{static}                                                   
\end{figure}

For what concerns the instantaneous response it gives a constant contribution to the Fourier transform of the third-order current. For a field applied at an arbitrary angle $\theta$ its  polarization dependence reads:
\bea
\lb{jinstpll}
J_{Inst,\pll}/A_0^3&=&\frac{T^0_x}{24}[\cos^4\theta+\sin^4\theta]\\
J_{Inst,\perp}/A_0^3&=&\frac{T^0_x}{96} \sin(4\theta),
\eea
where $T^0_x$ is the average kinetic energy along the $x$ direction. For a given disorder level $J_{Ints,\pll}$ has then a shallow minimum at $\theta=\pi/4$, and in the homogeneous case it is of the same order of the diamagnetic response at $\pi/4$, as one can see by comparing Fig.\ \ref{static}b and Fig.\ \ref{jdiav0}b. As disorder increases the kinetic energy has negligible variation, see Fig.\ \ref{static}a, so overall the instantaneous term remains of the same order, resulting in a largely subleading contribution to the THG in the presence of disorder.

\section{Estimate of the scattering rate}\label{appTau}
Our estimate of the scattering rate is based on the spectral-weight conservation for the optical conductivity, which can be expressed for a lattice system as\cite{scalapino_prb93}
\begin{equation}\label{eq:d1}
\int_{0}^\infty d\omega \sigma'_{s/n}(\omega)=-\frac{\pi}{2} T^0_\alpha
\end{equation}
where $\alpha$ indicates the direction of the applied vector potential and $T^0_\alpha$ is the kinetic energy along $\alpha$. In the parabolic-band approximation $\langle t_\alpha\rangle$ simply reduces to $n/m$ and one recovers the so-called f-sum rule.
When the system enters the SC state all the spectral weight which disappears at finite frequency must be recollected in the superfluid delta-like response at $\omega=0$, in order to still satisfy Eq.\ \pref{eq:d1}. We can then separate in full generality the SC optical conductivity into the delta-function
contribution, determined by the superfluid stiffness $D_s$, and the
regular part\cite{scalapino_prb93,cea_prb14}
\begin{equation}\label{eq:d2}
  \sigma'_s(\omega)=\pi D_s\delta(\omega)+\sigma'_{s,reg}(\omega)
\end{equation}
whereas the normal state is described by the Drude formula
\begin{equation}
\lb{drude}
  \sigma'_n(\omega)=D_c \frac{\tau}{1+\omega^2\tau^2}
\end{equation}
with the scattering time $\tau$ and the charge stiffness $D_c=\langle t_\alpha\rangle=0.781 t$ for the present doping level $n=0.875$.

As a first step we consider the case of weak disorder where
all the contribution to $D_s$ stems from the gap region
with $\sigma'_s(0<\omega\le 2\Delta)=0$.

Since we work at zero temperature from Eq. (\ref{eq:d1}) one finds 
\begin{eqnarray}
  \int_{2\Delta}^\infty d\omega \sigma'_{s}(\omega)&=&\int_{2\Delta}^\infty d\omega \sigma'_{n}(\omega) \\
  \frac{\pi}{2}D_s &=& D_c \mbox{atan}(2\Delta\tau) \label{eq:dd2}
\end{eqnarray}
which results in the following equation for the scattering strength
\begin{equation}\label{eq:d3}
  2\Delta \tau=\tan(\frac{\pi}{2}\frac{D_s}{D_c}) \,.
\end{equation}

Thus, Eq. (\ref{eq:d3}) allows for the estimate of the scattering
rate once the superfluid stiffness $D_s$ is known. For the parameter values and disorder levels considered here $D_s$ has been computed in previous work, see Ref.s \cite{cea_prb14,seibold_prb15}. For the clean system, where $D_s=D_c$, Eq. (\ref{eq:d3}) yields $\tau\to \infty$.

Clearly, for arbitrary disorder the delta-function in $\sigma'_{s}(\omega)$
also gets weight
from excitations above the gap. Then Eq. (\ref{eq:dd2}) has to be  generalized
to
\begin{equation}\label{eq:d5}
  \frac{D_s}{D_c}=\frac{2}{\pi}\mbox{atan}(2\Delta\tau)
  +\frac{2}{\pi D_c}\int_{2\Delta}^{\infty}\left\lbrack
  \sigma'_n(\omega)-\sigma'_s(\omega)\right\rbrack
\end{equation}
where the integral on the {\it r.h.s.} also depends on $\tau$
via $\sigma'_{n,s}(\omega)$.

\begin{figure}[htb]
  \includegraphics[width=7.5cm,clip=true]{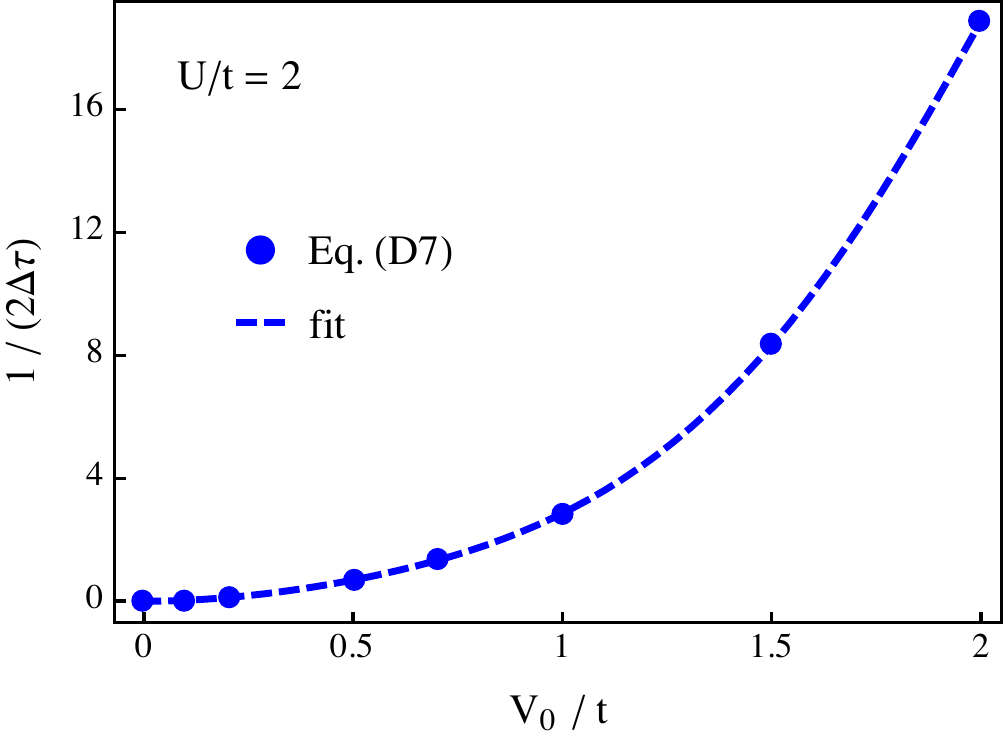}
  \caption{Estimate of the transport scattering rate $\tau$ as a function of $V_0$. The data points denote the values of $V_0/t$ reported in Table \ref{tabtau}. The solid line is a polynomial fit, that has been used to infer $\tau$ for different values of $V_0$ investigated in the present work.} 
\label{figtau}                                                   
\end{figure}

In order to evaluate Eq. (\ref{eq:d5}) we have adopted an expression
for $\sigma'_{s}(\omega)$ \cite{Zimmermann} which is valid for arbitrary
disorder and thus can be used to estimate $2\Delta\tau$ for a given
disorder strength $V_0/t$.
Corresponding results are summarized in Table I and shown in Fig.\ \ref{figtau}.

\begin{table}[htb]
\label{tabtau}
  \begin{tabular}{c|c|c|c|c}
    $V_0/t$  &$\Delta$ & $D_s/t$ & $2\Delta\tau$ &$1/(2\Delta\tau)$\\
\hline\hline
    $0.1$ & $0.359$ & $0.76$ & $28.7$ & 0.035 \\
    $0.2$ & $0.343$ & $0.714$ & $8.4$ & 0.12 \\
    $0.5$ & $0.3003$ & $0.51$ & $1.46$ & 0.68\\
    $0.7$ & $0.28$ & $0.39$ & $0.75$ & 1.33 \\
    $1.0$ & $0.243$ & $0.255$ & $0.35$ & 2.86 \\
    $1.5$ & $0.21$ & $0.12$ & $0.12$ & 8.33 \\
    $2.0$ & $0.24$ & $0.06$ & $0.053$ & 18.87
  \end{tabular}
  \caption{Disorder strength $V_0/t$ (as specified below Eq. 4), half of the spectral gap $\Delta$,
    superfluid stiffness $D_s$ and
  resulting scattering parameter $2\Delta\tau$ as obtained from Eq.\ (\ref{eq:d5}).}
\end{table}

\bibliography{Literature.bib}

\end{document}